\documentclass[pre,showpacs,amssymb]{revtex4}

\usepackage{epsfig}
\usepackage{wasysym}
\usepackage{ae}
\usepackage[T1]{fontenc}
\usepackage{bbold,psfrag}

\newcommand{\di}{\displaystyle}
\newcommand{\mc}{\mathcal}
\newcommand{\be}{\begin{equation}}
\newcommand{\ee}{\end{equation}}
\newcommand{\bd}{\begin{displaymath}}
\newcommand{\ed}{\begin{displaymath}}
\newcommand{\bea}{\begin{eqnarray}}
\newcommand{\eea}{\end{eqnarray}}
\newcommand{\wh}{\widehat}

\def\verylongrightarrow{\relbar\joinrel\relbar\joinrel\relbar\joinrel\rightarrow}

\begin{document}

\title{The Importance of Boundary Conditions
for Fluctuation Induced Forces between Colloids at Interfaces}

\date{\today}

\author{Hartwig Lehle$^{1,2}$}
\author{Martin Oettel$^{1,2,3}$}

\affiliation{$^1$ Max-Planck-Institut f\"ur Metallforschung, Heisenbergstr. 3,
  D-70569 Stuttgart, Germany, }
\affiliation{$^2$ Institut f\"ur Theoretische und Angewandte Physik, Universit\"at Stuttgart, 
             Pfaffenwaldring 57, D-70569 Stuttgart, Germany}
\affiliation{$^3$ Institut f\"ur Physik, WA 331, Johannes-Gutenberg-Universit\"at
  Mainz,
55099 Mainz, Germany}


\begin{abstract}
We calculate the effective fluctuation induced
force between spherical or disk-like
colloids trapped at a flat, fluid interface
mediated by thermally excited capillary waves. 
This Casimir type force is determined by 
the partition function of the system which in turn is calculated in a functional integral
approach, where the restrictions on the capillary waves
imposed by the colloids
are incorporated by auxiliary fields.
In the long-range regime
the fluctuation induced force is shown to depend 
sensitively on the boundary conditions imposed 
at the three-phase contact line between the colloids and the two fluid phases.
The splitting of the fluctuating capillary wave field into a mean-field
and a fluctuation part leads to competing repulsive
and attractive contributions, respectively, which give
rise to  cancellations of the leading terms.
In a second approach based on multipole expansion of the
Casimir interaction,
these cancellations can be understood from the vanishing of certain
multipole moments enforced by the boundary conditions.
We also discuss the connection of the different types of boundary conditions
to certain external fields acting on the colloids which appear to be
realizable by experimental techniques such as the laser tweezer method. 
\end{abstract}



\maketitle

%

\section{Introduction}\label{sec:intro}
The structure formation of nanoscopic
colloidal particles  
adsorbed at fluid interfaces has attracted an increasing interest in recent
 years because of the various applications such systems exhibit, as in the
design of nanoscale devices \cite{Joann01} ({\em e.g.} for optical applications),
and also because of the physical insights offered, {\em e.g.}  for the understanding 
of protein aggregation on cell membranes.
The practical importance arises from the fact that
 nano- and microcolloids  are very effectively trapped by fluid interfaces \cite{Pier80}. 
The high stability of partially wetting colloids (with their size ranging from  nm to 
$\mu$m) at interfaces
enables the formation of two-dimensional ordered structures
and also of rather complex mesoscale patterns
({\em cf.} Refs.~\cite{Ghez97,Ghez01,Gar98,Gar98a,Stam00,Ques01,Sea99,Che05,Niko02}).
In spite of the numerous  experimental and theoretical efforts
in the last decade,
the nature of the effective forces between the colloids 
governing the arrangement of the colloids 
trapped at the interface is not fully understood yet. This holds in particular for
the mesoscale pattern formation which points to sizeable and quite long--ranged 
attractions between the colloids.
For the colloidal interactions at interfaces
three different regimes may be distinguished.
For colloid sizes of about $0.5-5\,\mu {\rm m}$,
capillary forces are important.
Though gravity is known to be unimportant in this
regime, electrostatic forces (either caused only by the charge distributions
on the colloid surface and the fluid phases, or additionally imposed
by an external field) 
may lead to interfacial deformations
and, hence, to considerable capillary interactions \cite{Wuerg05,Oet05,Oet05b}.
In this regime, also strong effects of colloid surface inhomogeneities
leading to an undulated three--phae contact line appear to be relevant
\cite{Stam00}.
In the opposite limit of colloid sizes of a few nanometers,
effects of interparticle correlations within the interface become important
and determine the structural properties of the system.
In this regime the colloidal interactions should  be treated 
by truly microscopic fluid theories like
density functional or inhomogeneous integral equation 
approaches (for a general scheme, see Ref.~\cite{Oet05a}, and for
results on correlations within a free interface, see Refs.~\cite{Mill86,And05}). 
In between these two regimes, the microscopic one which must be tackled with
the full power of classical statistical mechanics and the macroscopic one
where thermal fluctuations appear to be unimportant, a coarse--grained 
picture of the fluctuating fluid interface should be applied. Within such a picture,   
the properties of fluid interfaces are
very well described by an effective capillary wave Hamiltonian
which governs both the equilibrium interface configuration
and the thermal fluctuations (capillary waves) around this equilibrium 
(or mean-field) position.
As postulated by the
Goldstone theorem the capillary waves are long-range correlated.
The interface breaks the continuous translational symmetry
of the system, and in the limit of vanishing external fields -- like gravity --
it has to be accompanied by
easily excitable long wavelength (Goldstone) modes -- precisely
the capillary waves \cite{Jas86}.
The fluctuation spectrum of the capillary waves will be modified by colloids
trapped at the interface and therefore leads to fluctuation--induced forces
between them.
In the case of anisotropic colloids (rods) these forces 
have been evaluated in Ref.~\cite{Gol96} and shown to lead to an orientational dependence.
Furthermore we note that there is numerous work on the force between inclusions
on membranes where the membrane shape fluctuations take the role of capillary waves 
, see {\em e.g.} Refs.~\cite{Goul93, Gol96}.  
In this paper, we shall address 
the fluctuation--induced effects of capillary waves  
on rotationally symmetric colloids, 
i.e., spherical or disk-like ones,
and pay special attention to the effects of the boundary conditions at the
three--phase contact line (where the fluid interface meets the colloid surface)
on the large--distance behavior of the fluctuation--induced force.
In previous work \cite{Leh06} we have established the independence of the 
short--distance behavior on the boundary conditions. For colloids at contact,
the fluctuation--induced force is attractive and diverging, albeit the
divergence is somewhat slower
than for the ubiquitous {\it van der Waals (vdW)}   forces. However, as discussed
in Ref.~\cite{Leh06}, the fluctation--induced force dominates for specially prepared system
(such as index--matched ones) and hence it leads to colloidal aggregation at the interface.     


Put in broader context,
the prototype of such a fluctuation induced force as considered here is the original
Casimir effect of the zero-temperature, long-ranged quantum fluctuations
of the electromagnetic field confined between two conducting plates
\cite{Bor01} that was predicted by Hendrik Casimir in 1948.
In 1978 this was extended by Fisher and de Gennes \cite{Fish78}
to critical systems like bulk fluids near the critical point or in binary mixtures
near their critical demixing point
where instead of long-ranged quantum fluctuations
one has long-ranged critical fluctuations \cite{Kre94}.
An even more common
example for long-range correlated fluctuations in
thermal systems are the already mentioned  Goldstone modes accompanying
broken continuous symmetries. 
These modes are ubiqitous in condensed matter physics 
with famous examples being 
director fluctuations in liquid crystals \cite{Ajd91}, phase fluctuations
in superfluids \cite{kard92} or 
  capillary waves 
at fluid interfaces mentioned above \cite{Wein89}.


Both, the great experimental progress (see {\em e.g.} Ref.~\cite{Lam97}) and the fundamental 
interest in the
Casimir effect have  triggered numerous studies in this field in the last twenty
years. In spite of this effort, still the only exactly solvable case
is Casimir's original one of two parallel infinite planes. 
An almost closed solution has been obtained 
for the plate-cylinder case very recently \cite{Emig06}.
In more complex geometries, 
analytical results usually are only available in
asymptotic ranges, 
 and the strong dependence
of the Casimir force on the geometrical arrangement of the system
has been shown 
in the case of the fluctuating medium being a free scalar field
for  various configurations
like, e.g., immersed
metallic spheres \cite{Jaf04} or cylinders \cite{Emig06} and plates.
A standard, though uncontrolled, approximation for the Casimir interaction
between one curved body and a plate or two curved bodies at short distances is the well-known
 proximity force approximation 
which presumes pairwise additivity and
which is in the same spirit as the Derjaguin approximation \cite{Der34} well known
from calculations
of the van der Waals force between macroscopic bodies.  
The limitations of this approach
are, e.g., investigated  perturbatively and numerically
in Refs.~\cite{Bue04, gies06}.

In a study on the pairwise addititvity of the Casimir force \cite{Gol00}
Golestanian investigated the force between spheres in a fluctuating medium
described by a free scalar field in arbitrary dimension $D$. Using a terminology
borrowed from electrostatics, the scalar ``field'' corresponds to the
potential and if the spheres are presumed to be metallic, one can distinguish two
types of boundary condition: {\em (i)} the spheres are grounded, corresponding
to zero potential at their surface (Dirichlet boundary conditions) and {\em (ii)}
the spheres are isolated with constant charge, in which case one averages over the
surface potential. The asymptotic Casimir force between two spheres at distance
$d$ strongly depends on the type of boundary condition; in case {\em (i)}
the author finds $F_{\rm ground}\sim 1/d^{2(D-2)+1}$ and for case
{\em (ii)} $F_{\rm iso}\sim 1/d^{2D+1}$. In relation to this study, our present
work is only concerned with the case $D=2$. The boundary 
conditions introduced above correspond to the physically
realizable conditions of a fixed colloid with pinned three--phase contact line
(grounded case) and a verticallly fluctuating colloid with pinned contact line
(isolated case). Furthermore we will identify two more physically realizable
boundary conditions (arbitrarily fluctuating colloid with pinned contact line and
colloid with freely fluctuating contact line) which lead to different results   
for the Casimir force again. Also we will find that in $D=2$ the general result
for $F_{\rm ground}$ is modified since care is needed to perform the limit
to a massless fluctuating field.
We will elucidate the appearance of the
leading terms in an asymptotic expansion of the
Casimir force from two different computational schemes.
In the first one, the force is calculated
by  splitting   the fluctuating field
into a mean field and a fluctuation part
as, {\em e.g.}, it is well known from the functional integrals
in quantum mechanics. 
Then the partition function consists of two parts
which can be computed separately
and lead to a repulsive and an
attractive contribution to the Casimir interaction, respectively.
The leading non-vanishing term in the
asymptotic behaviour for large colloid separations
then strongly depends on the type of boundary
conditions under consideration.
In an alternative approach 
this can be understood  
from an auxiliary multipole expansion of the interaction between
the colloids similar as in Ref.~\cite{Gol00}.
One can
relate each type of boundary conditions 
to a suppression of 
certain auxiliary multipoles on the colloids and 
the leading term of the Casimir force is obtained from
the first non-vanishing multipole-multipole interaction.


The paper is organized as follows. In Sec.~\ref{sec:model}
we will introduce the model system by an effective Hamiltonian.
We will show how the partition function can be calculated 
via functional integrals in two distinct ways mentioned above
and how the various boundary conditions can be implemented.
Then we will compute the corresponding functional
integrals in Secs.~\ref{sec:longrange}, \ref{sec:derja} and \ref{numerics}
in the asymptotic regimes of long and short colloid separations $d$
and numerically in the full range of $d$, respectively,
and compare analytical with numerical
results in the discussion part, Sec.~\ref{sec:numres}.
In Sec.~\ref{sec:numres}, we will also discuss 
possible experimental tests for this type of Casimir force
through the implications
of an additional external potential for the colloids,
which may, e.g., be realized by  optical tweezers.

%
%
\section{Model}\label{sec:model}
\begin{figure}
 \psfrag{g}{\footnotesize $\gamma$}
 \psfrag{g1}{\footnotesize $\gamma_{\rm I}$}
 \psfrag{g2}{\footnotesize $\gamma_{\rm II}$}
\psfrag{r0}{\footnotesize $r_0$}
  \psfrag{I}{\footnotesize I}
   \psfrag{II}{\footnotesize II}
   \psfrag{d}{\footnotesize $d$}
   \psfrag{R}{\footnotesize $R$}
   \psfrag{h}{\footnotesize $h$}
   \psfrag{q}{\footnotesize $\theta$}
   \psfrag{z=0}{\footnotesize $z=0$}
\includegraphics[width=0.6\textwidth]{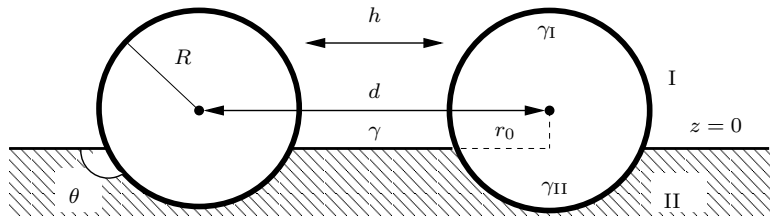}
\caption{Side view on the reference configuration (here the colloids are assumed to be
spheres). 
}
\label{fig1}
\end{figure}

\begin{figure}
\psfrag{x}{$x$}
\psfrag{y}{$y$}
\psfrag{d}{$d$}
\psfrag{R}{$r_0$}
\psfrag{DS1}{$\partial S_{1,\rm ref}$}
\psfrag{DS2}{$\partial S_{2,\rm ref}$}
\psfrag{S1}{$S_{1,\rm ref}$}
\psfrag{S2}{$S_{2,\rm ref}$}
\psfrag{r1}{$r_1$}
\psfrag{p1}{$\varphi_1$}
\psfrag{r2}{$r_2$}
\psfrag{p2}{$\varphi_2$}
\includegraphics[width=0.9\textwidth]{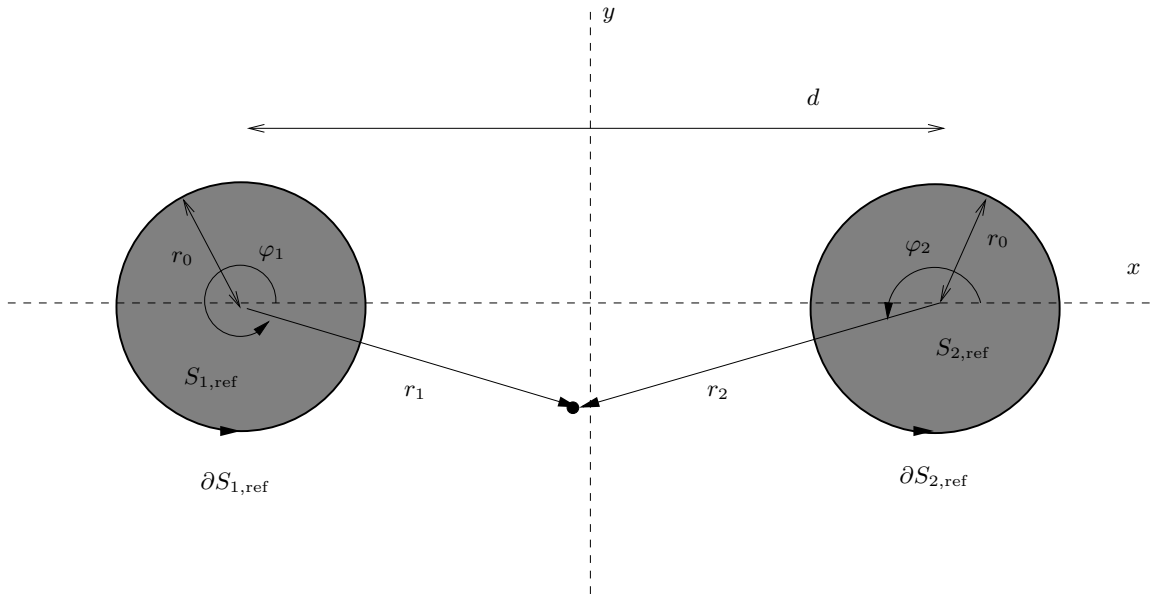}
\caption{
Reference configuration with two circles $S_{i,\rm ref}$ representing
the reference contact line on the colloid surfaces (here the colloids may be
disks or spheres). 
}
\label{fig2}
\end{figure}

In this section  we derive an effective Hamiltonian
for the free energy changes associated with
thermally excited height fluctuations of
the interface between two fluid phases I and II
at which two nano-- or microscopic colloids are trapped.
As described above, this configuration is very stable
against thermal fluctuations for colloids with a partially
wetting surface. 
The colloids are assumed to be 
either spherical with radius $R$ or disklike with radius $R$ and thickness $H$.
In the absence of charges and for
 colloid sizes $R\alt 1\,\mu$m
 the weight of the particles can be neglected.
Thus the equilibrium configuration of minimal
free energy is the flat interface and spherical colloids are positioned 
 such that Young's law holds at the horizontal 
three-phase contact line which is a circle with radius $r_0 = R \sin\theta$.
Young's angle, measured through phase II (assumed to be of higher density than phase 
I), 
is determined by
$\cos\theta =(\gamma_{\rm I}-\gamma_{\rm II})/\gamma$ where $\gamma$ is the surface
tension of the interface between I/II, and $\gamma_{{\rm I}\,[{\rm II}]}$ is the surface
tension between the colloid and phase I [II], respectively
(c.f. fig. \ref{fig1}). 
For disklike colloids, the contact line is either the
upper ($\theta<\pi/2$) or lower ($\theta>\pi/2$) circular edge, so that 
its radius is given by $r_0 = R$.
For both spheres and disks the cross-section of the
colloids with the flat interface is given by the interior of a circle
which we refer to as $S_{i,\rm ref}$ below, and,
therefore both cases can be treated within the same model in the following.
We take the flat interface $S_{\rm men,ref}$ as the reference configuration  with respect to
which free energy changes are measured, and choose it to be the
plane $z=0$ with the two circular holes $S_{i,{\rm ref}}$
of radius $r_0$ in it: $S_{\rm men,ref}=\mathbb{R}^2\setminus \cup_iS_{i,\rm
  ref}$ 
 (see fig. \ref{fig2}).
Deviations from the planar reference interface $z=0$ are assumed to be small
and without overhangs or bubbles.
That allows
using the Monge representation $(x,y,z=u(x,y))=({\bf x},z=u({\bf x}))$ as a
parametrization of the actual interface positions.
The free energy costs for thermal fluctuations 
around the flat reference interface
is determined by the the change in interfacial 
energy of {\em all} interfaces (I/II, colloid/I and colloid/II):
\be\label{Htot1}
\mc{H}_{\rm tot}=\gamma\Delta A_{\rm men}
+\gamma_{\rm I}\Delta A_{\rm I}
+\gamma_{\rm II}\Delta A_{\rm II}
\;.
\ee
The first term in Eq.~(\ref{Htot1})
expresses the energy needed for
creating the additional meniscus area associated with the
height fluctuations and is given by 
\bea\label{ham2}
\gamma\Delta A_{\rm men}
&=&
\gamma \int_{S_{\rm men}} d^2x\,\sqrt{1+ (\nabla u)^2}
-\gamma \int_{S_{\rm men,ref}} d^2x
\nonumber\\
&\approx&
\frac{\gamma}{2} \int_{S_{\rm men,ref}} d^2x\,  (\nabla u)^2
+\gamma \Delta A_{\rm proj} \,.
\eea 
In Eq.~(\ref{ham2}), $S_{\rm men}$ is the meniscus area projected onto the plane $z=0$
(where the reference interface is located)
and $S_{\rm men,ref}$ is the meniscus in the reference configuration mentioned
above. $\Delta A_{\rm proj}= \int_{S_{\rm men}\setminus 
S_{\rm men,ref}}\!\!\!\!\!\!\!\!d^2x$ is the change
in projected meniscus area with respect to the reference configuration.
In the second line we have applied a small gradient expansion
which is valid for slopes $|\nabla u|\ll 1$ and which provides the long wavelength
description of the interface fluctuations we are interested in. 
Note that the first line of Eq.~(\ref{ham2})
constitutes the  drumhead model which is well-known in the renormalistion group
analysis of interface problems, but is also used for 
the description of elastic surfaces 
(c.f. Ref.~\cite{Jas86}).

As discussed in the Introduction, 
the center of colloid $i$ may fluctuate vertically (measured by $h_i$)
around the reference position
as well as the contact line itself may do.
This leads 
to changes in the interfacial areas colloid/I[II] 
($\Delta A_{\rm I[II]}\neq 0$ in Eq.~(\ref{Htot1})).
In order to determine the corresponding  energy costs 
we introduce the vertical position 
of the contact line at colloid $i$ as a function of the polar angle $\varphi_i$, defined
on the reference contact line circles $\partial S_{i,{\rm ref}}$,
by its Fourier expansion
\be\label{conlin}
f_i = 
u(\partial S_{i,\rm ref})=\sum_{m=-\infty}^\infty  P_{im}\,e^{im\varphi_i}
\;
\ee
and refer to 
the Fourier coefficients $P_{im}$ as 
boundary multipole
moments below.
Since $f_i$ is real, we have $P_{im}=P_{i{-m}}^*$. 
Following Ref.~\cite{Oet05}, the free energy changes associated with
$\Delta A_{\rm proj}$ and $\Delta A_{\rm I[II]}$ can be expanded
up to second order in $h_i$  and $f_i$, and may therefore be collected
in a boundary Hamiltonian ${\cal H}_{i,\rm b}$ (see App.~\ref{app:Hb}):
%
\begin{eqnarray}
\label{Hb}
\mc{H}_{{\rm b},i}[f_i,h_i]
&=&
\gamma \Delta A_{\rm proj}
+\gamma_{\rm I}\Delta A_{\rm I}
+\gamma_{\rm II}\Delta A_{\rm II}
\nonumber\\
&=&
\frac{\pi\gamma}{2}\left[ 2(P_{i0}-h_i)^2+ 4\sum_{m\ge 1} 
|P_{im}|^2\right]\;.
\end{eqnarray}
%
Putting Eqs.~(\ref{Htot1}),(\ref{ham2}) and (\ref{Hb}) together, 
the total free energy change is the sum 
\bea\label{Htot2}
\mc{H}_{\rm tot}
&=&
\mc{H}_{\rm cw}+\mc{H}_{\rm b,1}+\mc{H}_{\rm b,2}
=
\frac{\gamma}{2} \int_{S_{\rm men,ref}} d^2x\,  (\nabla u)^2
+\mc{H}_{\rm b,1}+\mc{H}_{\rm b,2}
\eea
of the capillary wave Hamiltonian $\mc{H}_{\rm cw}$
which describes the free energy differences
associated with the additional interfacial area
created by the height fluctuations, 
and the boundary Hamiltonians $\mc{H}_{\rm b,i}$
combining all effects related to fluctutations of
the three-phase contact line on the colloid surfaces.
As is well-known, the
 Hamiltonian $\mc{H}_{\rm cw}$ is plagued with
both a short-wavelength and a long-wavelength
divergence which, however,  can be treated by physical cutoffs.
The natural
short-wavelength cut-off is set by
the molecular length-scale $\sigma$ of the
fluid at which the capillary wave model
ceases to remain valid. 
The long wavelength divergence is reminiscent to the fact
that the capillary waves are Goldstone modes. 
Of course, in real systems 
the gravitational field provides a natural damping for capillary waves.
Accounting also for the costs  in gravitational energy associated 
with the interface height fluctuations,
therefore, introduces a long wavelength cutoff
and leads to an additional term (``mass term'') in the capillary wave Hamiltonian,
\bea\label{capwav}
\mc{H}_{\rm cw}
&=&
\frac{\gamma}{2} \int_{S_{\rm men,ref}} d^2x\,  
\left[(\nabla u)^2+\frac{u^2}{\lambda_c^2}
\right]
\eea
 Here the capillary length is given by 
 $\lambda_c = [\gamma/(|\rho_{\rm II}-\rho_{\rm I} |\, g)]^{1/2}$, where $\rho_i$
 is the mass density in phase $i$ and $g$ is the gravitational constant. 
 Usually, in simple liquids, $\lambda_c$ is in the range of millimeters
and, therefore, is by far the longest length scale in the
system. In fact, here it plays the role of a long wavelength cutoff
of the capillary wave Hamiltonian $\mc{H}_{\rm cw}$, 
and we will discuss our results in the limit  $\lambda_c \gg R$ 
and $\lambda_c \gg d$.
However, as we will see below,
care is required when taking
 the limit $\lambda_c \to \infty$ (corresponding to $g \to 0$), since 
logarithmic divergencies appear \cite{Saf94}.
Another common way to introduce a long-wavelength cut-off
is the finite size $L$ of any real system.
As discussed in Ref.~\cite{Oet05}, the precise way of incorporating
the long-wavelength cut-off is unimportant for the effects on the colloidal
length scale. As an example, 
in both approaches the
width of the interface related to
the capillary wave is logarithmically divergent,
$\langle u(0)^2\rangle \sim \ln \lambda_c[L]/\sigma $.


%
Via the integration
domain of $\mc{H}_{\rm cw}$, 
the total Hamiltonian  of the system, Eq.~(\ref{Htot2}), implicitly  depends
on the geometric configuration.
This leads to a free energy which depends on the
mean distance $d$ between the colloid centers
and gives rise to an effective force $F(d)=-(\partial \mc{F})/(\partial d)$ 
as function of the mean local distance
between the colloid centers which is determined by 
 the free energy
${\mathcal F}(d)= -k_{\rm B} T \ln {\mathcal Z}(d)$. The partition function
${\mathcal Z}(d)$ is obtained by a functional integral over all possible interface 
configurations $u$ and $f_i$; the boundary configurations are included by 
$\delta$-function constraints,  
\begin{equation}\label{Z1}
\mathcal{Z}=\mathcal{Z}_0^{-1} \int \mathcal{D}u\,
\exp \left\{-\frac{\mathcal{H}_{\rm cw}[u] }{k_{\rm B}T} \right\}\;
\prod_{\rm i=1}^2 
\int \mathcal{D}f_i
\prod_{{\bf x}_i \in \partial S_{i,\rm ref}}
\delta [u({\bf x}_{ i})-f_{ i}({\bf x}_{ i})]
\exp \left\{-\frac{\mathcal{H}_{{\rm b},i}[f_i,h_i]}{k_{\rm B}T} \right\}
\;.
\end{equation}
Here $\mathcal{Z}_0$ is a normalization factor such that
$\mathcal{Z}(d\to\infty)=1$
and ensures a proper regularization of
the functional integral.
Via the $\delta$-functions the interface field $u$ is
coupled to the contact line height $f_i$ and therefore,
the boundary Hamiltonians $\mc{H}_{i,\rm b}$
have a crucial influence on the resulting effective
interaction between the colloids as we will see below.
In the next subsection we discuss 
possible situations for the boundary conditions
at the three phase contact line and specify the
corresponding integration measure $\mc{D}f_i$.
\subsection{Boundary conditions at the three-phase contact line}
\label{sec:modelbc}
%
We shall discuss two different realizations of the boundary conditions for
the contact line, namely 
\begin{itemize}
\item Case (A): The contact lines and the vertical colloid positions fluctuate
freely;
this corresponds to the physical situation of smooth, spherical colloids.
In this case, the integration measure is given by $\mc{D}f_i=  \int dh_i\prod_m dP_{
im}$ 
and encompasses integration over {\em all} boundary multipole moments.
\item Case (B): The contact lines (the circles $\partial S_{i,{\rm ref}}$ in the
reference configuration) are pinned to the colloid surface.
This corresponds to disklike colloids or Janus spheres consisting of two different
materials, or to colloids with a very rough surface.
Within the pinning case, we furthermore distinguish the following three situations:
\begin{itemize}
 \item[(B1)] The colloid positions are frozen, thus there are no
integrations over the boundary terms. 
 \item[(B2)] The vertical positions of the colloids 
fluctuate freely, thus boundary monopoles must be included in the integration measure 
 so that $\mc{D}f_i= dh_i\, dP_{i0}\,\delta(P_{i0}-h_i)$.
 \item[(B3)] The vertical position and the orientation of the colloids (tilts) 
fluctuate freely. Up to second order in the tilts this corresponds
to the inclusion of boundary dipoles in the integration measure, thus
$\mc{D}f_i=dh_i\, dP_{i0}\, dP_{i1}\,
dP_{i{-1}}\,\delta(P_{i0}-h_i)$.
\footnote{
Since $P_{im}=P_{i{-m}}^*$, real and imaginary parts
of $P_{i{\pm m}}$ are not independent of 
each other, and therefor we define the measure
$dP_{im}\,dP_{i{-m}} \equiv  d{\rm Re}P_{im}\,d{\rm Im}P_{i{-m}}$.
} 
\end{itemize}
The $\delta$-function
expresses the pinning condition.
\end{itemize}

In the following subsections we introduce two distinct schemes to compute
the partition function $\mc{Z}(d)$ in Eq.~(\ref{Z1}) which
 allows us to discuss the results from different
perspectives. The first one is based
on a splitting of $\mc{Z}$ in a mean-field and a fluctuation part,
whereas for the second one we extend 
the fluid interface -- artificially -- to the interior
of the reference circles $S_{i,\rm ref}$.
As we will see, in the first approach, the mean-field part
leads to a repulsive and the fluctuation part to an attractive
Casimir force. The interplay of these two contributions 
gives rise to an cancellation of the leading terms
in the asymptotic range $r_0 \ll d$ which sensitively depends
on the chosen boundary conditions.
In the second approach, the asymptotic Casimir force can be expanded
in multipole interactions of auxiliary fields defined on $S_{i,\rm ref}$ and 
the appearance of a different behaviour in the long-range
regime for different boundary conditions can be understood straightforwardly
as the result of vanishing 
multipole moments of the auxiliary fields, enforced by the
boundary conditions.
 
%
%
\subsection{Mean-field and fluctuation part}\label{subs:mf}
As the capillary wave Hamiltonian is Gaussian in the
field $u$ of the local interface position, a 
standard procedure for the evaluation of the 
functional integral (\ref{Z1})
is the decomposition into 
a mean-field and a fluctuation part, 
\be\label{mffluc}
u=u_{\rm mf}+v\;.
\ee
The mean-field part solves the Euler--Lagrange equation of the
capillary wave Hamiltonian $\mc{H}_{\rm cw}$,
\be\label{eleq}
 (-\Delta+\lambda_c^{-2})\,u_{\rm mf}=0
\ee
 with the boundary condition
$u_{\rm mf}\,|_{ \partial S_{i,{\rm ref}}}=f_{ i} $. 
Consequently the fluctuation part
is pinned to zero at the contact line,
$v\,|_{ \partial S_{i,{\rm ref}}}=0$.
Then the partition sum  $\mathcal{Z}=\mathcal{Z}_{\rm fluc}\mathcal{Z}_{\rm mf}$ 
separates into a product of a fluctuation part 
independent of
the boundary conditions, and a  mean field part 
depending on the boundary conditions (which may fluctuate theirselves, see the cases 
(A), (B2), and (B3)): 
%
\begin{eqnarray}\label{Zsep}
 \mathcal{Z}_{\rm fluc} &=& \mathcal{Z}_0^{-1} \int \mathcal{D}v\,
\prod_{\rm i=1}^2 \prod_{{\bf x}_i \in \partial S_{i,{\rm ref}}}\delta ( v({\bf x}_i))
\exp \left\{-\frac{\mathcal{H}_{\rm cw}[v]}{k_{\rm B}T} \right\}\;,  \nonumber \\
 \mathcal{Z}_{\rm mf} &=& \prod_{\rm i=1}^2
\int \mathcal{D}f_i
\, \exp\left\{ -\frac{\gamma}{2 k_{\rm B}T} \sum_{i}
\int_{\partial S_{i,\rm ref}}d\ell_i f_i({\bf x_i})\,
( \partial_n u_{\rm mf}(\bf{x}_i))\right\}
\exp \left\{-\frac{\mathcal{H}_{{\rm b},i}[f_i,h_i]}{k_{\rm B}T} \right\}
 \;.
\end{eqnarray}
The first exponential in $\mathcal{Z}_{\rm mf} $ stems from applying Gauss' theorem
to the energy associated with $u_{\rm mf}$. In this term
$ \partial_n u_{\rm mf}$ denotes the normal derivative of the mean--field
solution towards the interior of the circle $\partial S_{i,\rm ref}$. 

For intermediate asymptotic distances between the colloids 
($r_0 \ll d \ll \lambda_c$), $\mathcal{Z}_{\rm fluc}$
and $ \mathcal{Z}_{\rm mf}$ are analyzed separately in Secs.~\ref{sec:fluc} and
\ref{sec:mf}, respectively.

This procedure to split the functional integral into a mean field part and 
a fluctuation part with a zero boundary condition is equivalent to the
standard calculation of the quantum mechanical path integral for a single particle
in a harmonic potential. The mean field part corresponds to the classical action
and the fluctuation part evaluated becomes the fluctuation determinant
(see, e.g., Ref.~\cite{kleinert95}).

{
\psfrag{+}{+}
\psfrag{=}{=}
\psfrag{fixed}{fixed}
\psfrag{mean field}{mean--field}
}

\subsection{Alternative approach (Kardar's method)}\label{subsec:kardarmodel}
An alternative scheme to
calculate the partition function of the two colloids and the
fluctuating interface without splitting it into a mean-field and
fluctuation part can be devised if  
the interface height field $u(x,y)$ which enters the functional integral
for ${\cal Z}$ is extended
to the interior of the 
circles $S_{i,\rm ref}$. Thus the measure of the functional integral for ${\cal Z}$ is  
extended by ${\cal D}u({\bf x})|_{{\bf x} \in S_{i,\rm ref}}$ and the
integration domain in the capillary wave Hamiltonian is enlarged to
encompass the whole $\mathbb{R}^2$.
We note that physically the free energy of the system
must not change since in the interior of $S_{i,\rm ref}$ the interface is pinned
to the colloid surface. 
A method similar to this ansatz was introduced in Ref.~\cite{Gol96} by
Kardar {\em et al.} which
investigates effective forces between rods on 
fluctuating membranes and films, 
and therefore we refer to it as Kardar's method in the following. 

On the colloid surfaces, the interface height field is given by 
the three phase contact line, $u(\partial S_{i,\rm ref}) \equiv f_i$.
We extend $u$ continuously to the interior of the circles $S_{i,\rm ref}$
via  
\be\label{artcoll}
u(S_{i,\rm ref})
\equiv
f_{i,\rm ext}(r_i,\varphi_i)
=
\sum_{m} \left(\frac{r_i}{r_0}\right)^{|m|} \,
 P_{im}e^{i m\varphi_i} \;,
\ee
where $r_i$ and $\varphi_i$ are the polar coordinates with
respect to  
circle $S_{i,\rm ref}$
and where the boundary multipoles satisfies $P_{im}=P_{i{-m}}^*$. 
Note that the choice of
$u(S_{i,\rm ref})$ is not unique
since the continuity at the boundaries $\partial S_{i,\rm ref}$
is the only requirement. 
The specific choice in Eq.~(\ref{artcoll}) is
convenient for the further calculations since $\Delta f_{i,\rm ext}=0$
in $S_{i,\rm ref}\backslash \partial S_{i,\rm ref} $.
Extending the integration domain of the capillary wave Hamiltonian in Eq.~(\ref{Htot2}),
$\Omega=\mathbb{R}^2 \setminus\cup_i S_{i,\rm ref} \to \mathbb{R}^2$
generates an additional energy contribution
\bea\label{Hcorr1}
-\mc{H}_{i,\rm corr}
&=&
\frac{\gamma}{2}\int_{S_{i,\rm ref}}d^2x\,
\left[(\nabla u)^2+\frac{u^2}{\lambda_c^2}
\right]
\nonumber\\
&\simeq& 
2\pi\gamma\sum_{m\ge 1}m\,|P_{im}|^2\;.
\eea
In Eq.~(\ref{Hcorr1}) we have already omitted the contributions from the
gravitational term in $\mc{H}_{\rm cw}$ which are of order $(r_0/\lambda)^2 \ll 1$.
From Eq.~(\ref{Hcorr1}) we see that 
the additional terms created by the extension of the integration domain of the
capillary wave Hamiltonian 
$\mc{H}_{\rm cw}$  are {\it not} constant in
the -- possibly fluctuating -- boundary multipole moments $P_{im}$  and
therefore lead to artificial contributions to the
partition 
function $\mc{Z}$. These unphysical contributions 
have to be corrected by adding $\mc{H}_{i,\rm corr }$ to the 
extended capillary wave Hamiltonian
 $\mc{H}_{\rm cw}[\Omega \equiv \mathbb{R}^2]$. 
The total Hamiltonian then reads
\bea\label{Htotkard}
\mc{H}_{\rm tot}
&=&
\mc{H}_{\rm cw}+\sum_{i=1}^2[\mc{H}_{i,\rm b}+\mc{H}_{i,\rm corr}]\;.
\eea
As in the previous sections the 
partition function is written as a functional integral
over all possible configurations of the 
interface position $u$ and the boundary lines, expressed by $f_i$,
\begin{equation}\label{Zkard1}
\mathcal{Z}=\mathcal{Z}_0^{-1} \int \mathcal{D}u\,
\prod_{\rm i=1}^2 
\int \mathcal{D}f_i
\prod_{{\bf x}_i \in  S_{i,\rm ref}}
\delta [u({\bf x}_{ i})-f_{i,\rm ext}({\bf x}_{ i})]
\exp \left\{-\frac{\mathcal{H}_{{\rm tot}}[f_i,u]}{k_{\rm B}T} \right\}
\;,
\end{equation}
where the product over the $\delta$-functions
enforces the pinning of the interface at the
positions of the colloids.
In contrast to Eq.~(\ref{Z1}), these product extends over all
${\bf x} \in S_{i,\rm ref}$ instead of $\partial S_{i,\rm ref}$, only.
The analysis of ${\cal Z}$ in this form for intermediate asymptotic distances 
$d$ is presented below
in Sec.~\ref{sec:Kardar}.

\section{Long-range behavior}\label{sec:longrange}
In this section analytical expressions for the
fluctuation induced force
in the intermediate asymptotic regime $r_0\ll d \ll \lambda_c$
are calculated.
In Secs.~\ref{sec:fluc} and \ref{sec:mf} we will evaluate the
partition functions of
fluctuation and the mean-field part, respectively.
From these two contributions we obtain the asymptotic
form of the total Casimir force, which in turn  will be 
calculated directly in Sec.~\ref{sec:Kardar} applying
the approach introduced in Sec.~\ref{subsec:kardarmodel}. 
\subsection{Fluctuation part}\label{sec:fluc}
The fluctuation part appears in all cases (A) and (B1)--(B3) for the
boundary conditions introduced in Sec.~\ref{sec:modelbc}. 
In the case
(B1) (pinned contact line with frozen colloid) 
it constitutes the full result for the partition function because $u_{\rm mf}=0$
and $\mc{Z}_{\rm mf}=1$.
We express the $\delta$-functions 
in the fluctuation part of the partition function (\ref{Zsep})
by  their integral representation via
 auxiliary fields $\psi_i ({\bf x}_i)$ defined on the
interface boundaries $\partial S_{i,\rm ref}$. This enables us to integrate out 
the field $u$ leading to
\begin{eqnarray}
\label{Zaux}
 \mathcal{Z}_{\rm fluc} =
\int \prod_{i=1}^2 \mc{D}\psi_i\,
\exp\left\{
-\frac{k_{\rm B}T}{2\gamma}\sum_{i,j=1}^2
\int_{\partial S_{i,\rm ref}}d\ell_i \int_{\partial S_{j,\rm ref}}d\ell_j\,
\psi_i ({\bf x}_i)\,G(|{\bf x}_i-{\bf x}_j|)\,\psi_j({\bf x}_j)\right\}\;,
\end{eqnarray} 
where $d\ell_i$ is the infinitesimal line segment on the circles $\partial S_{i,\rm ref}$.
In Eq.~(\ref{Zaux}) we introduced the Greens function of the operator 
$(-\Delta + \lambda_c^{-2})$
which is given by
$G({\bf x})=K_0(|{\bf x}|/\lambda_{ c})/(2\pi)$ where $K_0$ is the modified Bessel
function of the second kind.
 In the range $d/\lambda_c \ll 1$ and $r_0/\lambda_c \ll 1$,
we can use 
 the asymptotic form of the $K_0$ for small arguments, such that
$2\pi\,G(|{\bf x}|) \approx -\ln(\gamma_{\rm e} |{\bf x}|/\,2\lambda_c) $. Here,
$\gamma_{\rm e} \approx 1.781972$  is the Euler-Mascheroni constant exponentiated.

We introduce auxiliary multipole moments as the
Fourier-transforms of 
the auxiliary fields $\psi_i$ on the contact line circles $\partial S_{i,\rm ref}$,
\bea\label{auxmomdef}
\wh{\psi}_{im}
&=&
\frac{r_0}{2\pi}
\int_0^{2\pi}
d\varphi_i\,e^{{\rm i}m\varphi_i}\,\psi_{i}({\bf x}_i(\varphi_i))
\eea
The analogous multipole decomposition for the Greens function
$G(|{\bf x}_i-{\bf x}_j|)$ is calculated in App. \ref{app:schwing}.
Using it, 
the double integral in the
exponent of Eq.~(\ref{Zaux}) can be written as a double sum
over the Fourier components, consisting of a self-energy part
%
\bea\label{flucse}
\mc{G}_{\rm self}
&=&\sum_i \int_{\partial S_{i,\rm ref}}d\ell_i \int_{\partial S_{i,\rm ref}}d\ell_i\,
 \psi_i ({\bf x}_i)\,G(|{\bf x}_i-{\bf x}_i|)\,\psi_i({\bf x}_i) 
\nonumber\\
&=&
\sum_i\left[
-2\pi\,\ln\left(\frac{\gamma_{\rm e} r_0}{2\lambda_c}\right)
|\wh{\psi}_{i0}|^2
+\sum_{n> 0}\frac{2\pi}{n}
|\wh{\psi}_{in}|^2
\right]\;,
\eea
and an interaction part 
\bea\label{auxmom2}
\mc{G}_{\rm int}
&=&
2 \int_{\partial S_{1,\rm ref}}d\ell_1 \int_{\partial S_{2,\rm ref}}d\ell_2\,
 \psi_1 ({\bf x}_1)\,G(|{\bf x}_1-{\bf x}_2|)\,\psi_2({\bf x}_2) 
\nonumber\\
&=&
2\pi
\left[
-2\ln\left(\frac{\gamma_{\rm e}d}{2\lambda_c}\right)\,
\wh{\psi}_{10}\wh{\psi}_{20}
\right.\nonumber\\&&\quad\left.
+
 \sum_{{m,n=0} \atop{m+n\geq 1} }
\frac{(-1)^{n}}{m+n}\left({m+n}\atop n\right)\left(\frac{r_0}{d}\right)^{m+n}
\,
\left(\wh{\psi}_{1m}\wh{\psi}_{2n}+\wh{\psi}_{1{-m}}\wh{\psi}_{2{-n}}
\right)
\right]\;.
\eea
Inserting these expressions into Eq.~(\ref{Zaux}), 
the functional integral over the auxiliary fields can be written
as an integral over their multipole moments.
The partition function then reads
\begin{eqnarray}\label{Zauxmom}
  \mc{Z}_{\rm fluc}
  &=&
  \int \prod_{i=1}^2 \mc{D}\psi_i\,
  \exp\left\{-
    \frac{k_{\rm B}T}{2\gamma}
    \left(
      \begin{array}{c}
        {\bf\widehat{\Psi}}_1\\
        {\bf\widehat{\Psi}}_2
      \end{array}
    \right)
    ^{\rm T}
    \left(\begin{array}{cc}
        \wh{ \bf G}_{\rm self} & \wh{\bf{ G}}_{\rm int} \\
        \wh{\bf{G}}_{\rm int} & \wh{\bf{G} }_{\rm self}
      \end{array}\right)
    \left(
      \begin{array}{c}
        {\bf \widehat{\Psi}}_1\\
        {\bf \widehat{\Psi}}_2
 \end{array}
\right)
\right\}\;,
\end{eqnarray}
where the vectors 
${\bf  \widehat{\Psi}}_i=(\wh{\psi}_{i0},\wh{\psi}_{i1},\wh{\psi}_{i {-1}},\dots)$
contain the auxiliary multipole moments of colloid $i$.
The coupling matrix ${\bf \wh{G}}$ which contains the 
Fourier modes of the Greens function $G({\bf x}_i-{\bf x}_j)$
has a block structure. The self energy
submatrices $\bf{\wh{G}}_{\rm self}$ which describes
the coupling between auxiliary moments of the same colloid
are diagonal, and their elements can be read off Eq.~(\ref{flucse}). 
The offdiagonal blocks $\bf{\wh{G}}_{\rm int}$ characterise
the interaction between the multipole moments residing on different
colloids and are assigned by Eq.~(\ref{auxmom2}). Note that
all matrix elements coupling modes with positive and negative
$m$ vanish.
From Eq.~(\ref{Zauxmom}) we find that
the 
fluctuation part of the free energy reads
\bea\label{Fflucaux}
\mc{F}_{\rm fluc}
=-k_{\rm B}T\ln Z_{\rm fluc}
\propto
k_{\rm B}T\ln \det{\widehat{\bf G}} \;,
\eea
where 
in principle the matrix $\wh{\bf G}$ is  infinite dimensional and
divergent and therefore requires regularisation.
However, after performing the logarithm to calculate the free energy
these  divergencies only reside in terms independent
of the geometrical arrangement of the two colloids, and therefore
the derivative of the logarithm with respect to $d$ which corresponds to the
Casimir force is finite.
Hence, the physically important properties of $\mc{Z}_{\rm fluc}$
reside in the offdiagonal blocks of the matrix $\bf{G}$
which describe the interaction between the  colloids  depending on
 their separation $d$, whereas  the self energy part 
ensures the correct normalisation of the interaction.
Eq.~(\ref{auxmom2}) allows for a systematic expansion of
the logarithm in Eq.~(\ref{Fflucaux}) 
in powers of $r_0/d$,
\bea\label{Fflucexp}
\mc{F}_{\rm fluc}(d)
&=&
\frac{k_{\rm B}T}{2}
\sum_{n}f_{2n}^{\rm fluc}
\,\left(\frac{r_0}{d}\right)^{2n}\;,
\eea
where
the coefficients $f_{2n}^{\rm fluc}$ depend 
on the logarithms
$-\ln(\gamma_{\rm  e}d/\,2\lambda_c))$ and $-\ln(\gamma_{\rm e}r_0/\,2\lambda_c)$.
We show the first few coefficients in Tab. \ref{coefftable}.
 The number of auxiliary multipoles included in
the calculation of the asymptotic form of
$\mc{F}_{\rm fluc}$ in Eq.~(\ref{Fflucexp})
is determined by the desired order in $r_0/d$. 
The individual contributions in this expansion can be understood
as (possibly higher order) auxiliary multipole-multipole interactions,
related to specific products of the matrix elements of $\bf{G}$.
So, e.g., all the logarithmic contributions to the coefficients in 
Eq.~(\ref{Fflucexp}) are related to interactions of the auxiliary monopoles.
Note that we cannot perform 
$\lim_{\lambda_c \to \infty}f_{2n}^{\rm fluc}$
because of the logarithmic contributions mentioned above.
These logarithmic divergencies are reminiscent of the 
long-range correlations of the capillary
waves which lead to a width of the free interface which diverges 
logarithmically with $\lambda_c$
or with the system size if gravity is absent.
For the Casimir force itself, however,
we find a finite value in the limit $\lambda_c\to \infty$.
In the asymptotic range  $r_0/d \ll 1$ and
in the limit $\lambda_c/d \to \infty$ 
the leading term of the fluctuation force  
is governed by the first term in the series Eq.~(\ref{Fflucexp}), 
$f_{0}^{\rm fluc}$,
and reads
\begin{equation}
\label{ffluc}
F_{\rm fluc}=k_{\rm B}T \frac{\partial}{\partial d}\ln \mathcal{Z}_{\rm fluc}
  \;\to\;-\frac{k_{\rm B}T}{2}\frac{1}{d\ln (d/r_0)} + \mc{O}(d^{-3})\;,
\qquad \frac{d}{r_0}\gg 1,\;\frac{d}{\lambda_c}\to 0\;.
\end{equation}
Note, however, that the limit $\lambda_c \to \infty$
here is slowly converging with a leading correction term of the order
$1/(d\ln\lambda_c)$,
\be\label{limlam}
-\frac{\partial f_0^{\rm fluc}}{\partial d}
\qquad \stackrel{d/\lambda_c \ll 1}{\verylongrightarrow }\qquad
-\frac{1}{d\ln (d/r_0)}
\left[
1+\frac{3\ln(r_0/d)}{2\ln(r_0/\lambda_c)}
+\mc{O}\big((\ln(r_0/d)^2/\ln(r_0/\lambda_c)^2\big)
\right]
\ee
and that the free energy difference corresponding to Eq.(\ref{ffluc}),
${\cal F}(d) \sim \ln\ln( r_0/d)$, is actually ill--defined, 
because the effective colloidal interaction in case (B1) -- fixed
colloids and pinned interface -- is only meaningful for a finite capillary
length 
$\lambda_c$.
\begin{table}
\caption{Leading coefficients of Eq.~(\ref{Fflucexp})}
\label{coefftable}
\renewcommand{\baselinestretch}{2.5}
\begin{tabular}{|c|p{0.9\textwidth}|}
\hline
&\\
2n & \hspace*{3cm} $2 f_{2n}^{\rm fluc}/k_{\rm B}T$ 
\\ & 
\\
\hline\hline
0 
& 
\bea\label{ffluc0}
\ln\left[\ln\left(\frac{\gamma_{\rm e}r_0}{2\lambda_c}\right)^2-\ln\left(\frac{\gamma_{\rm
    e}d}{2\lambda_c}\right)^2\right]
\equiv\ln g_0^{\rm fluc}
\nonumber
\eea 
\\
\hline
2
&
\bea\label{ffluc2}
\frac{2\ln\left(\frac{\gamma_{\rm e}d}
{2\lambda_c}\right)}{g_0^{\rm fluc}}
\nonumber\eea
\\
\hline
4 
&  
\bea\label{ffluc4} 
 -2
+\frac{1
+\ln\left(\frac{\gamma_{\rm e}r_0}{2\lambda_c}\right)
+\ln\left(\frac{\gamma_{\rm e}d}{2\lambda_c}\right)
}{g_0^{\rm fluc}}
-\textstyle{\frac{1}{2}}\left(f_{2}^{\rm fluc}\right)^2
 \nonumber
\eea  
\\
\hline
6
&
\bea\label{ffluc6}
-8
+\frac{1
+\textstyle{\frac{2}{3}}\ln\left(\frac{\gamma_{\rm e}r_0}{2\lambda_c}\right)
+4\ln\left(\frac{\gamma_{\rm e}d}{2\lambda_c}\right)
}{g_0^{\rm fluc}}
-f_2^{\rm fluc}f_4^{\rm fluc}
+\textstyle{\frac{1}{3}}\left(f_{2}^{\rm fluc}\right)^3
\nonumber\eea
\\
\hline
8
&
\bea\label{ffluc8}
-30
+\frac{\textstyle{\frac{11}{12}}
+\textstyle{\frac{1}{2}}\ln\left(\frac{\gamma_{\rm e}r_0}{2\lambda_c}\right)
+7\ln\left(\frac{\gamma_{\rm e}d}{2\lambda_c}\right)
}{g_0^{\rm fluc}}
-\textstyle{\frac{1}{2}}\left(f_{4}^{\rm fluc}\right)^2
-f_2^{\rm fluc}f_6^{\rm fluc}
+
\left(f_{2}^{\rm fluc}\right)^2f_4^{\rm fluc}
-\textstyle{\frac{1}{4}}\left(f_{2}^{\rm fluc}\right)^4
\nonumber\eea
\\
\hline
\end{tabular}
\end{table}
\subsection{Mean-field part}\label{sec:mf}
The calculation of  $\mc{Z}_{\rm mf}$ (Eq.~(\ref{Zsep})) requires 
the solution of the Euler-Lagrange equation~(\ref{eleq}),  
%
$(-\Delta+\lambda_c^{-2})\,u_{\rm mf}({\bf x})=0$
for  ${\bf x} \in \mathbb{R}^2\setminus \cup_i S_{i,\rm ref}$
and the (fluctuating) boundary conditions 
\be \label{mfeq}
u_{\rm mf}({\bf x}_i)=f_i({\bf x}_i)
\ee
with ${\bf x}_i \in \partial S_{i,\rm ref}$ 
and $u_{\rm mf}({\bf x})\to 0$ 
for $|{\bf x}|\to \infty$.
In fact a solution to a similar problem (with Neumann boundary conditions) was given
in Ref.~\cite{kral91} in terms of bipolar
coordinates  which, however, is involved if applied to 
the general Dirichlet boundary conditions in our case.
For our purpose it is more convenient to write the 
solution as a superposition 
\bea\label{superpos}
u_{\rm mf}({\bf x}) &=& 
u_1({\bf x}-{\bf r}_1)+u_2({\bf x}-{\bf r}_2) 
\equiv
u_1(r_1,\varphi_1)+u_2(r_2,\varphi_2) \;, 
\eea
where ${\bf r}_i$ is the center position  of  circle $S_{i,\rm ref}$.
The  general solutions $u_i$ of the mean-field equation~(\ref{eleq}) 
in polar coordinates $(r_i,\varphi_i)$ with respect to the centers of the reference
contact line circles $S_{i,\rm ref}$ (see Fig.~\ref{fig2} ) 
can be written as linear combinations
\bea\label{superpos2}
u_i(r_i,\varphi_i) 
&=& 
\sum_m 
A_{im}\,a_{im}(r_i,\varphi_i) 
\;\label{superpos3}
\eea
The functions $a_{im}$ are defined by
\bea\label{mfeigenf}
a_{i0}(r_i,\varphi_i)&=& \frac{ K_0 ( r_i/\lambda_{ c})}{K_0 ( r_0/\lambda_{ c})} 
\approx
\frac{\ln(\gamma_{\rm e}r_i/2\lambda_c)}{\ln(\gamma_{\rm e}r_0/2\lambda_c)}
\nonumber\\
a_{im}(r_i,\varphi_i) &=&  \frac{ K_m ( r_i/\lambda_{ c})}{K_m ( r_0/\lambda_{ c})}\, e^{i m\varphi_i} 
\approx \left(\frac{r_0}{r_i}\right)^{|m|}e^{i m\varphi_i} 
\;,
\eea
which are normalized for convenience
and where 
we have used the asymptotic form of the
modified Bessel functions $K_m$ for small arguments $r_i/\lambda_c \ll 1$.
As the solution has to match the boundary conditions 
at both circles $\partial S_{1,\rm ref}$ and $\partial S_{2,\rm ref}$,
we project $u_{\rm mf}$ onto 
the complete set of functions on $\partial S_{1,\rm ref}$, $e^{i m\varphi_1}$, 
and on $\partial S_{2,\rm ref}$, $e^{i m\varphi_2}$, respectively.
The expansion coefficients of these projections of $u_{\rm mf}$ must equal
 the boundary multipole moments at the corresponding circle which leads to a
system of linear equations for the expansion coefficients
$\{A_{im},B_{im}\}$ of the $u_i$,
\bea\label{proj}
A_{1m}+\sum_{n=-n_{\rm max}}^{n_{\rm max}}
a_{1,mn}\, A_{2n}
&=&P_{1m}
\nonumber\\
A_{2m}+\sum_{n=-n_{\rm max}}^{n_{\rm max}}
a_{2,mn}\, A_{1n}
&=&P_{2m}
\;.
\eea
In Eq.~(\ref{proj}),
the matrix coefficients $a_{1,mn}$ are
given by the projection coefficients 
\bea\label{projcoeff}
a_{1,mn}
&=&
\frac{1}{2\pi}\int_0^{2\pi}
d\varphi_1\,
e^{i m\varphi_1}\,a_{2n}(r_2,\varphi_2)\;,
\eea
 and analogously $a_{2,mn}$ is given in terms of the projection  of 
 $a_{1m}(r_1,\varphi_1)$ onto 
 $e^{i m\varphi_2}$. 
 In Eq.~(\ref{proj}), we have truncated the 
Fourier expansions at a maximum
 order $n_{\rm max}$. 
Indeed, the numerical solution shows rapid convergence of the expansions
in Eq.~(\ref{superpos2})
in the whole range of colloid-colloid separations $d$, such that they can
be truncated  at $n_{\rm max}\approx 20$.
Through the inversion of the linear system~(\ref{proj}) 
the coefficients $A_{im}$ and
$B_{im}$ can be expressed as linear combinations of the boundary multipoles,
%
\bea\label{lincomb}
A_{im}
&=&
\sum_{j=1}^2\sum_{n=-n_{\rm max}}^{n_{\rm max}}
 p_{imjn} P_{jn} 
\;.
\eea
Using the asymptotic forms of the functions $a_{im}$ given
in Eq.~(\ref{mfeigenf}), expressing $r_2=r_2(d,r_1,\varphi_1)$
and $\varphi_2=\varphi_2(d,r_1,\varphi_1)$
by the polar coordinates with respect to circle $S_{1,\rm ref}$
and vice versa for circle $S_{2,\rm ref}$, and
expanding the projection coefficients $a_{i,mn}$ in Eq.~(\ref{proj})
in $r_0/d$, the coefficients $p_{imjn}$ in Eq.~(\ref{lincomb}) can be 
written as a power series in $r_0/d$.
Inserting the solution (\ref{superpos}) with 
the expression (\ref{lincomb}) for the $A_{im}$  
into
the line integrals for
the mean field energy 
$\mc{H}[u_{\rm mf}]$ in the exponent of Eq.~(\ref{Zsep}),
we can write $\mc{H}[u_{\rm mf}]$ 
as a quadratic form of the boundary multipole moments $P_{im}$,  
 described by a matrix ${\bf E}$ with block structure,
\bea\label{mfe}
\mc{H}_{\rm mf}
&=& 
-\frac{\gamma r_0}{2}\sum_i
 \int_{0}^{2\pi}d\varphi_i\,
 f_i(\varphi_i)
\frac{
 \partial 
u_{\rm mf}({\bf x}(\varphi_i))
}{\partial r_i}
 +\sum_i\mc{H}_{{\rm b},i}[f_i,h_i]
\nonumber\\ 
&=&
 \frac{\gamma}{2}
 \left(
 \begin{array}{c}
 {\bf\hat{f}}_1\\
 {\bf\hat{f}}_2
 \end{array}
 \right)
 ^{\rm T}
 \left(\begin{array}{cc}
 {\bf E}_{\rm 1\,self} & {\bf E}_{\rm int} \\
 {\bf E}_{\rm int} & {\bf E}_{\rm 2\, self}
 \end{array}\right)
 \left(
 \begin{array}{c}
 {\bf \hat{f}}_1\\
 {\bf \hat{f}}_2
 \end{array}
 \right)
+\pi\gamma\sum_i (P_{i0}-h_i)^2
\;,
\eea
Here, the submatrices ${\bf E}_{i,\rm self}$ and ${\bf E}_{\rm int}$  
describe the coupling energy of the multipole moments of the same contact line 
$f_i$ and from different contact lines, respectivly, and their elements are
given by
\bea\label{mfself}
E_{i,\rm self}^{mn}
&=&
2\pi\delta_{mn}(1-\delta_{m0})-\int_{0}^{2\pi} r_0 d\varphi_i \,e^{im\varphi_i}\,
\sum_{k}
p_{ikin}\frac{ \partial}{\partial r_i} a_{ik}(r_i,\varphi_i)\;, 
\\
E_{12,\rm int}^{mn}
&=&
-\int_{0}^{2\pi}r_0d\varphi_1 \,e^{im\varphi_1}
\sum_k
p_{2k1n}
\frac{ \partial}{\partial r_1} a_{2k}(r_2(\varphi_1),\varphi_2(\varphi_1)\;,
\label{mfint}
\eea
and analogously for $E_{21,\rm int}^{mn}$.
Applying again the expansions of the functions $a_{ik}$ in powers
of $r_0/d$ in the expression for $E_{12,\rm int}^{mn}$
and using the series expressions for the coefficients $p_{imjn}$,
the matrix elements can be written as a power series in
$r_0/d$ in the asymptotic range $r_0\ll d\ll \lambda_c$.
The mean field part of the partition sum Eq.~(\ref{Zsep}) 
can then be written as
\begin{equation}\label{Zmfcalc}
\mc{Z}_{\rm mf} = \int \prod_i\mc{D} f_i\; \exp\left \{ - \frac{\mc{H}_{\rm mf}}{k_{\rm
B} T}\right\}\;
\end{equation}
such that the $d$--dependent part of $\mc{Z}_{\rm mf}$ is proportional to 
$\det {\bf E}$.
As discussed in Sec.~\ref{sec:modelbc}, the functional measure 
is given by a product $\mc{D}f_i=dh_i\prod_{m=-M}^{M}dP_{im}$,
where $M$ depends on the choosen model for the boundary
condition at the contact line. 
Inserting the power series of the matrix elements $E_{i,\rm int}^{mn}$
and  $E_{i,\rm self}^{mn}$ and expanding 
 the free energy  
$\mc{F}_{\rm mf}(d) \propto k_{\rm B}T\ln\det{\bf E}$,
we arrive at a series similar in form to the fluctuation part,  Eq.~(\ref{Fflucexp}),
\bea\label{Fmfexp}
\mc{F}_{\rm mf}(d)
&=&
\frac{k_{\rm B}T}{2}
\sum_{n}f_{2n}^{\rm mf}
\,\left(\frac{r_0}{d}\right)^{2n}\;,
\eea
where again the coefficients $f_{2n}^{\rm mf}$
are functions of the monopole self-energy and
interaction coupling elements,
$-\ln(\gamma_{\rm e}r_0/2\lambda_c)$
and $-\ln(\gamma_{\rm  e}d/2\lambda_c)$, respectively.
The exact form of the $f_{2n}^{\rm mf}$ can be
recovered from Tab. \ref{coefftablemf}. 
Before discussing the analytic structure of the dependence on $d$, we remind the reader that
the integration measure $\mc{D} f_i$ differs between the cases
(A) -- no pinning, (B2) -- pinning and height fluctuation of the colloids,
    and (B3) -- pinning with collective height and tilt fluctuations of the
    contact line. 

In all these cases for the boundary conditions we have a mean-field contribution,
and the leading term  
of the free energy in the 
long-range regime $d\gg r_0$
(with $\lambda_c\to \infty$) is determined
by  the boundary monopole-monopole interaction 
and 
leads to a repulsive effective force between the colloids:
\begin{equation}
\label{fmf}
F_{\rm mf}=k_{\rm B}T \frac{\partial}{\partial d}\ln \mathcal{Z}_{\rm mf}
  \to \frac{k_{\rm B}T}{2}\frac{1}{d\ln (d/r_0)} +
  \mc{O}(d^{-3})\;,\qquad\frac{d}{r_0}\gg 1\;,\frac{d}{\lambda_c}\to 0\;.
\end{equation}
So the leading term of the mean-field contribution
is equal in size but opposite in sign to the fluctuation contributions,
and thus these  terms cancel. We will discuss the leading terms
of the total Casimir and their connection to the type of 
boundary conditions which is considered 
in the next subsection in more detail. 

\begin{table}
\caption{Leading coefficients $2 f_{2n}^{\rm mf}/k_{\rm B}T$ of
  Eq.~(\ref{Fmfexp})
for various boundary conditions.
The coefficients $f_n^{\rm fluc}$ are given in Tab.~\ref{coefftable}.
}
\label{coefftablemf}
\renewcommand{\baselinestretch}{2.5}
\begin{tabular}{|c|p{0.3\textwidth}|p{0.3\textwidth}|p{0.3\textwidth}|}
\hline
2n & case (B2): only monopoles  & case (B3): mono- and dipoles  & case (A): all multipoles
\\
\hline\hline
0 
& 
$-f_{0}^{\rm fluc}$
&
$-f_{0}^{\rm fluc}$ 
&
$-f_{0}^{\rm fluc}$
\\
\hline
2
&
$-f_{2}^{\rm fluc}$
&
$-f_{2}^{\rm fluc}$ 
&
$-f_{2}^{\rm fluc}$
\\
\hline
4 
&  
$-f_{4}^{\rm fluc}-2$
&
$-f_{4}^{\rm fluc}$
&
$-f_{4}^{\rm fluc}$
\\
\hline
6
&
$-f_{6}^{\rm fluc}-8$
&
$-f_{6}^{\rm fluc}$
&
$-f_{6}^{\rm fluc}$
\\
\hline
8
&
$-f_{8}^{\rm fluc}-31$
&
$-f_{8}^{\rm fluc}-18$
&
$-f_{8}^{\rm fluc}-2$
\\
\hline
\end{tabular}
\end{table}
\subsection{Total Casimir force for $d\gg r_0$}\label{sec:anares}
The $1/d$ asymptotics for both the mean-field and the
fluctuation part at hand, 
we will discuss
the total effective force $F$,
which due to $\mc{Z}=\mc{Z}_{\rm fluc}\mc{Z}_{\rm mf}$
 is  obtained as $F=F_{\rm fluc}+F_{\rm mf}$.
As discussed before, the mean-field result depends on
the type of boundary conditions under consideration.
In all cases 
 the large $d$ expansion of the mean--field part
of the partition sum 
(with $\lambda_c\to \infty$) leads to a repulsive effective force between the
 colloids, whereas the fluctuation contribution is attractive.
So the total Casimir force between the colloids is determined by an
interesting interplay between the two--dimensional ``bulk'' fluctuations and 
one--dimensional boundary fluctuations influencing the two--dimensional ``bulk'' by a change
of the mean-field.

Combining the expansions of the 
mean-field and the fluctuation free energies, Eqs.~(\ref{Fmfexp})
and (\ref{Fflucexp}), the asymptotic form of
the total Casimir can be written as a power series in $r_0/d$,
\bea\label{totalforce}
F(d)
&=&
-\frac{k_{\rm B}T}{2}\frac{\partial}{\partial d}\sum_{n}
\left[f_{2n}^{\rm mf}+f_{2n}^{\rm fluc}\right]
\left(\frac{r_0}{d}\right)^{2n}\;.
\eea
For a pinned contact line and fixed colloids,
$f_{2n}^{\rm mf}=0$ for all $n$, and the full
 Casimir force is given by the
fluctuation part result Eq.~(\ref{ffluc}).
As we have seen above, for fluctuating boundary
conditions,
the leading terms of $F_{\rm mf}$ and
$F_{\rm fluc}$ (Eqs.~(\ref{fmf}) and (\ref{ffluc})) cancel.
The same holds for the first subleading terms, 
for which we find $f_2^{\rm fluc}=-f_2^{\rm mf}$
for all types of boundary conditions with a mean-field
contribution.
The form of $f_{4}^{\rm mf}$, however, is different for
the cases (B2), and (B3) and (A), respectively.
In the simplest case of a pinned contact line with
height (or monopole) fluctuations of the colloid, we find
$f_{4}^{\rm mf}+f_{4}^{\rm fluc}=-1$, such
that in case (B1) the leading term of the fluctuation-induced
force in the limit
$\frac{d}{r_0}\gg 1\;,\frac{d}{\lambda_c}\to 0$
is attractive and stemming from the fluctuation part,
and can written as
\bea\label{lrmon}
    F \to {\di-\frac{\di 4\, k_{\rm B}T}{\di r_0}\,\left(\frac{\di r_0}{\di d}\right)^5}\;.
\eea
For a pinned contact line with tilt and height fluctuations
(boundary mono- and dipoles, case (B3)) and for the generic case
of an unpinned and fluctuating contact line
(described by {\em all} boundary multipoles, case (A)),
we find that this term also cancels: $f_4^{\rm mf}=-f_4^{\rm mf}$
and $f_6^{\rm mf}=-f_6^{\rm mf}$. 
In these cases, the
first non-vanishing coefficient in the expansion Eq.~(\ref{totalforce})
is stemming from the fluctuation part as well. 
In case (B3) it is given by 
$f_{8}^{\rm mf}+f_{8}^{\rm fluc}=-9$
which leads to
\bea\label{lrdip}
 F \to {\di-\frac{\di 72\, k_{\rm B}T}{\di r_0}\,\left(\frac{\di r_0}{\di d}\right)^9}\;.
\eea
For case (A) we find $f_{8}^{\rm mf}+f_{8}^{\rm fluc}=-1$
 leading to an attractive Casimir force which
(in the limit $\frac{d}{r_0}\gg 1\;,\frac{d}{\lambda_c}\to 0$)
reads
\bea\label{lrA}
 F \to {\di-\frac{\di 8\, k_{\rm B}T}{\di r_0}\,\left(\frac{\di r_0}{\di d}\right)^9}\;.
\eea
\subsection{Kardar's method}\label{sec:Kardar}
Here we calculate the partition function directly
without splitting the fluctuating field $u$ into a mean-field and 
a fluctuation part.
The starting point is Eq.~(\ref{Zkard1}) from Sec.~\ref{subsec:kardarmodel}. 
The $\delta$-functions can again be expressed
by auxiliary fields $\psi_i$, now defined on the {\em two--dimensional} 
circular domains $S_{i,\rm ref}$
as opposed to the auxiliary fields of Sec.~\ref{sec:fluc} which are defined on the 
{\em one--dimensional} 
circles $\partial S_{i,\rm ref}$:
\begin{eqnarray}
\label{Zkard11}
 \mathcal{Z} &=&
\int \mc{D}u
\int \prod_{i=1}^2 \mc{D}\psi_i\int \mc{D}f_i\,
\exp\left\{
-\frac{\mathcal{H}_{{\rm tot}}[f_i,u]}{k_{\rm B}T}
+i\int_{S_i,\rm ref}d^2x\,
\psi_i({\bf x})[u({\bf x})-f_{i,\rm ext}({\bf x})]
\right\}\;.
\end{eqnarray} 
The function $f_{i,\rm ext}$ which describe the interface extension to
$S_{i,{\rm ref}}$ have been defined in Eq.~(\ref{artcoll}) in terms of a multipole
expansion in the $P_{im}$.
The total Hamiltonian contains the capillary wave Hamiltonian, the boundary and correction
terms (see Sec.~\ref{subsec:kardarmodel}) and reads 
\begin{equation}
 \mathcal{H}_{{\rm tot}} = 
\frac{\gamma}{2} \int_{\mathbb{R}^2} d^2x\,  
\left[(\nabla u)^2+\frac{u^2}{\lambda_c^2}\right] + \frac{\pi\gamma}{2}
 \sum_i\left[2(P_{i0}-h_i)^2+4\sum_{|m|\ge 1} (1-|m|)\,
 |P_{im}|^2 \right] \;.
\end{equation}
Similarly to the evaluation of the fluctuation part, Sec.~\ref{sec:fluc}, we introduce
multipole moments $\Psi_{im}$ of the auxiliary fields by inserting
unity into ${\cal Z}$, Eq.~(\ref{Zkard11}):
\bea\label{unity}
\mathbb{1}
&=&\int \prod_{i=1}^2
\prod_{m } d{\Psi}_{im} 
\,\delta
\left(\Psi_{im} -\int_{S_i}d^2x\,(r/r_0)^{|m|}e^{-im\varphi}\psi({\bf x})\right)
\;.
\eea
In contrast to the evaluation of the fluctuation term in Sec.~\ref{sec:fluc}, there
will be constraints on the lowest multipoles which contribute to ${\cal Z}$. 
To see this
we note that, after shifting $h_i \to h_i - P_{i0}$, the Hamiltonian 
$\mathcal{H}_{{\rm tot}}$ does not depend anymore on the boundary monopole moments
$P_{i0}$ and the dipole moments $P_{i1}$ and the only dependence of ${\cal Z}$ on these
moments is through the constraint function $f_{i,\rm ext}$. Recalling the definition of the
integration measure ${\cal D}f_i$ for the various boundary conditions, 
Sec.~\ref{sec:modelbc} and performing the integration over $P_{i0}$ and $P_{i1}$ 
where applicable, we immediately find
\begin{equation}
 \label{constraints}
  {\cal Z} \sim \left\{ 
   \begin{array}{ll} 
     {\displaystyle \int \prod_{i=1}^2 \prod_{m } d{\Psi}_{im} \dots \delta  
   ( \Psi_{i0} ) \dots} & \mbox{case (B2)} \\
     {\displaystyle \int \prod_{i=1}^2 \prod_m d{\Psi}_{im} \dots \delta ( \Psi_{i0})  \,
 \delta ( \Psi_{i-1})\, \delta ( \Psi_{i1}) \dots \qquad }& \mbox{cases (A) and (B3)}
   \end{array}
   \right.
\end{equation}
Having noticed these constraints on the auxiliary fields, we proceed by
integrating 
over the 
field $u$ in Eq.~(\ref{Zkard11}):
\begin{eqnarray}
\label{Zkard2}
 \mathcal{Z} &=&
\int \prod_{i=1}^2 \mc{D}\psi_i\int \mc{D}f_i\,
\exp\left\{
-\frac{k_{\rm B}T}{2\gamma}\sum_{i,j=1}^2
\int_{ S_{i,\rm ref}}d^2x_i \int_{ S_{j,\rm ref}}d^2x_j\,
\psi_i ({\bf x}_i)\,G(|{\bf x}_i-{\bf x}_j|)\,\psi_j({\bf x}_j)
\right. \nonumber\\ &&\left.
-\frac{\pi\gamma}{2k_{\rm B}T} \left[2(P_{i0}-h_i)^2+4\sum_{|m|\ge 1} (1-|m|)\,
 |P_{im}|^2 \right]
- i\sum_{i=1}^2\int_{S_{i,\rm ref}}{\rm d}^2x\,\psi({\bf x})f_{i,\rm ext}({\bf x})
\right\}\;,
\end{eqnarray} 
where -- as in Eq~(\ref{Zaux}) -- 
$G$ is the Greens function of the capillary wave Hamiltonian.
A somewhat longer calculation shows that ${\cal Z}$ can be split
into into an interaction part (coupling the auxiliary multipole moments
$\Psi_{im}$ for different colloid labels $i$), a self--energy part
(depending on $\Psi_{im}$ 
for each value of $i$ separately) and a remainder (the sum
of boundary and correction Hamiltonian):
\begin{eqnarray}
 \mathcal{Z} &=&  \int \prod_{i=1}^2 \prod_m d\Psi_{im} \int \mc{D}f_i\,
  \exp\left\{
-\frac{k_{\rm B}T}{2\gamma}\left(\mc{H}_{\rm int}[\Psi_{1m},\Psi_{2m}] +
  \mc{H}_{i,\rm self}[\Psi_{im}]\right) \right\} \times \nonumber \\
 \label{Zkard3} 
  &&\qquad \qquad 
 \exp\left( \frac{\pi\gamma}{2k_{\rm B}T}
 \left[2(P_{i0}-h_i)^2+\sum_{|m|\ge 1} (1-|m|)\,
 |P_{im}|^2 \right] -i \sum_m\Psi_{im} P_{im}
\right)
\end{eqnarray}
The interaction part
%
\bea\label{kardint}
\mc{H}_{\rm int}
&=&
2\int_{S_{1,\rm ref}}d^2x_1\int_{S_{2,\rm ref}}d^2x_2\,
\psi_1({\bf x}_1)G(|{\bf x}_1-{\bf x}_2|)\psi_2({\bf x}_2)
\nonumber\\
&=&
\frac{1}{2\pi}\left[
-2\ln\left(\frac{\gamma_{\rm e}d}{2\lambda_c}\right)\Psi_{10}\Psi_{20}
\right.\nonumber\\&&\left.
+\sum_{l_1,l_2=0 \atop l_1+l_2 \ge 0}
\frac{(-1)^{l_2}}{l_1+l_2}
{l_1+l_2 \choose l_2}
\left(\frac{r_0}{d}\right)^{l_1+l_2}
(\Psi_{1l_1}\Psi_{2l_2}+\Psi_{1{-l_1}}\Psi_{2{-l_2}})
\right]
\eea
is a bilinear form in the auxiliary multipole moments; it was derived 
using the multipole expansion of the Greens function
 $G(|{\bf x}_1-{\bf x}_2|) 
\simeq -\ln(\gamma_{\rm e}|{\bf x}_1-{\bf x}_2|/2\lambda_c)$ (valid for $d \gg r_0$)
which is
presented in App.~\ref{app:schwing} in more detail. The self--energy part
\begin{eqnarray}
\exp\left( -\frac{k_{\rm B}T}{2\gamma} \mc{H}_{i,\rm self}\right) &=&
 \int \prod_{i=1}^2 \mc{D}\psi_i\,\delta
\left(\Psi_{im} -\int_{S_i}d^2x\,(r/r_0)^{|m|}e^{-im\varphi}\psi({\bf x})\right)
 \exp\left( i\sum_m \Psi_{im}P_{im} \right)\times \\
 \nonumber && \qquad \exp\left(
-\frac{k_{\rm B}T}{2\gamma}
\int_{S_i}d^2x \int_{S_i}d^2x'\,
\psi_i({\bf x})\,G(|{\bf x}-{\bf x'}|)\,\psi_i({\bf x})
-i\int_{S_{i,\rm ref}}d^2x\,\psi_i({\bf x})\,f_{i,\rm ext}({\bf x})
\right)
\end{eqnarray}
is evaluated in App.~\ref{app:se}, with the result
\begin{equation}
\label{sekard}
  \mc{H}_{i,\rm self} = -|\Psi_{i0}|^2 \frac{\ln(\gamma_{\rm e} r_0/\,2\lambda_c)}{2\pi}
 + \sum_{m > 0} \frac{|\Psi_{im}|^2}{2\pi|m|}
\end{equation}
Combining Eqs.~(\ref{Zkard3}), (\ref{kardint}), and (\ref{sekard}), the partition
function can be written as
\begin{eqnarray}\label{Zkard6}
  \mc{Z}
  &=&
 \int \prod_{i=1}^2\prod_m
    \mc{D}\Psi_{im} \mc{D}f_i\,
  \exp\left\{-
    \frac{k_{\rm B}T}{2\gamma}
    \left(
      \begin{array}{c}
        {\bf{\Psi}}_1\\
        {\bf{\Psi}}_2
      \end{array}
    \right)
    ^\dagger
    \left(\begin{array}{cc}
        \wh{ \bf H}_{\rm self} & \wh{\bf{ H}}_{\rm int} \\
        \wh{\bf{H}}_{\rm int} & \wh{\bf{H} }_{\rm self}
      \end{array}\right)
    \left(
      \begin{array}{c}
        {\bf{\Psi}}_1\\
        {\bf{\Psi}}_2
 \end{array}
\right)
\right\}\;,
\end{eqnarray}
where the vectors 
${\bf{\Psi}}_i=
(\Psi_{i0},  P_{i0},
\Psi_{i1},P_{i1}, 
\Psi_{i{-1}},P_{i{-1}}
,\dots)$ -- in contrast to $\wh{\bf{\Psi}}_i$
in Sec.~\ref{sec:fluc} --
contain all involved auxiliary
and boundary multipole moments.
The elements of
 the matrix ${\bf{H}}$ describe the coupling of
these multipole moments, where the self-energy block
couples multipoles defined on the same circles $S_{i,\rm ref}$. 
The self energy matrix $\wh{ \bf H}_{\rm self}$ 
can be read off Eqs.~(\ref{Zkard3}) and (\ref{sekard}).
The elements of the interaction matrix $\wh{ \bf H}_{\rm int}$
are determined by the interaction energy $\mc{H}_{\rm int}$ in
Eq.~(\ref{kardint})
and couple the auxiliary multipole moments of
different colloids. All matrix elements
representing couplings of other multipoles are zero.

Similar as in Eqs.~(\ref{Zauxmom}) and (\ref{Zmfcalc}),
the exponent in Eq.~(\ref{Zkard6}) is a bilinear
form, however, here combined for all
types, boundary multipole moments $P_{im}$ and 
 auxiliary multipoles $\Psi_{im}$.
The computation of the partition function
amounts to the calculation of  $\det\wh{\bf H}$.
Expanding the logarithm of this
determinant for $r_0/d \ll 1$, and 
taking the derivative with respect to $d$,
we directly obtain the asymptotic form Eq.~(\ref{totalforce})
for the total Casimir force,
\bea\label{totalforce1}
F(d)
&=&
-\frac{k_{\rm B}T}{2}\frac{\partial}{\partial d}\sum_{n}
f_{2n}^{\rm kardar}
\left(\frac{r_0}{d}\right)^{2n}\;.
\eea
with $f_{2n}^{\rm kardar}=f_{2n}^{\rm fluc}+f_{2n}^{\rm mf}$ as it should be.
In contrast to the calculation before, the different leading power laws 
for the different cases (A) and (B1)--(B3)
can be understood easily. 
We note that the interaction between the auxiliary multipoles $\Psi_{1m}$ and $\Psi_{2n}$ 
in $\mc{H}_{\rm int}$, Eq.~(\ref{kardint}),
scales like $(r_0/d)^{m+n}$. Thus, the leading order of the total fluctuation induced
force between the two colloids is determined by the
first non-vanishing auxiliary multipole moment $\Psi_{im'}$ and (as follows
from $\det\wh{ \bf H}$) gives rise to a force $F(d) \propto 1/d^{2m'+1}$ (for $m'>0$)
or $F(d) \propto 1/(d\ln d)$ (for $m'=0$). 
As explained in the beginning of this subsection, the different
boundary conditions lead to certain constraints on the auxiliary multipoles:
According to Eq.~(\ref{constraints}), the leading term in $F(d)$
arises from a monopole-monopole interaction of the auxiliary field in case (B1), 
from a dipole-dipole interaction in case (B2),
and from a quadrupole-quadrupole interaction in cases (B3) and (A).
We remind the reader that
the constraints of vanishing auxiliary monopole and dipole moments  result from the 
independence of $\mc{H}_{\rm tot}$
of the boundary  monopole and dipole moments and that this is only captured correctly 
by the inclusion of
the correction Hamiltonian $\mc{H}_{\rm corr}$ (see Sec.~\ref{sec:Kardar}). 

At this point we insert the following observation: If the sum of boundary and correction Hamiltonian 
were zero, all multipole moments $\Psi_{im}$ would be zero and consequently all coefficients in the
expansion of the Casimir force in terms of $r_0/d$ would vanish -- i.e. the total Casimir force
would be zero. This happens for the boundary Hamiltonian \cite{Wie06}
\begin{eqnarray}
  {\cal H}_{{\rm b},i} &=& 2\pi\gamma \sum_{m\ge 1}m\,|P_{im}|^2  \\
    &=& \frac{\gamma}{16\pi} \int_{\partial S_{i,{\rm ref}}} \!\!\!\!\! d\phi
  \int_{\partial S_{i,{\rm ref}}} \!\!\!\!\!
  d\phi'\;\frac{ [f_i(\phi)-f_i(\phi')]^2}
   {\sin^2\left[ \frac{1}{2}(\phi-\phi') \right]}\;.
\end{eqnarray}
Thus we see that the boundary Hamiltonian needs to be of nonlocal nature in the contact line height 
$f_i$ to make the Casimir force vanish. 

%
%
 \section{Short--range behavior}\label{sec:derja}
 \subsection{Fluctuation part}\label{subsec:derja}
  In the
 opposite limit of small surface--to--surface distance $h=d-2r_0 \ll r_0$ the
 fluctuation force can be calculated by using
the Derjaguin (or proximity) approximation  \cite{Der34}.
 It consists in replacing the local
 force density on the contact lines by the  result for the fluctuation force
 per length $f_{\rm 2d}(\tilde h)$ between two parallel lines with
 a separation distance ${\tilde h}$ and integrating over the
 two opposite contact line half--circles to obtain the total effective force
 between the colloids.

The Casimir force between two parallel
surfaces was calculated in Ref.~\cite{Li91}
in a general approach and explicitly for
three-dimensional problems.
Applied to two dimensions
we obtain  $f_{\rm 2d}(\tilde h)= -k_{\rm B}T\,\pi/(24 {\tilde h}^2)$. 
Integrating over the opposing contact line half-circles
yields 
%
 \begin{eqnarray}
 \label{fflucderja}
 F_{\rm fluc} \approx  
 -2\frac{\pi k_{\rm
 B}T}{24}\int_0^{r_0}dy\,\frac{1}{\left(h+2r_0-2\sqrt{r_0^2-y^2}\right)^2}
 &\stackrel{r_0/h \to \infty}{\longrightarrow}&
 -k_{\rm B}T\, \frac{\pi^2}{48}\,
 \frac{r_0^{1/2}}{h^{3/2}}
 +\mc{O}(h^{-1/2})
  \;.
 \end{eqnarray}
for the dominant contribution
to the Casimir force from the fluctuation part
in the limit $h/r_0 \ll 1$.
Note, that this strong increase as $h\to 0$ is a consequence of the finite
 (mesoscopic) size of the colloids and is missed if the colloids are approximated by
 pointlike
 objects
 \cite{Kai05}. 

 \subsection{Mean-field part}
Here, we exemplify the asymptotic behaviour of the
mean--field force for close colloid separations $h \to 0$
by case (B2) for the boundary
conditions where only fluctuations of the boundary monopoles
occur. From the
numerical results (see next section) we observe that
the type of divergence of the mean--field force as $h \to 0$ is obtained correctly
by considering only monopole fluctuations; including higher multipole
moments affects the force only by a multiplicative constant.
(This is in marked contrast to the long-range regime.)

In order to apply the Derjaguin approximation, we calculate the mean-field
 between 
two parallel lines $\partial S_{1/2}$ on which the field is pinned to a
fluctuating value $u_{\rm mf}(\partial S_i)=P_{i0}$
(corresponding to monopole boundary conditions).
The mean-field
energy $\mc{H}_{\rm mf}$ in Eq.~(\ref{mfe})
is represented by a $2 \times 2$-matrix $\bf{E}$.
By diagonalizing $\bf{E}$ we can write
 $\mc{H}_{\rm mf}=\frac{\gamma\, L_y}{2}(e_{\rm s}Z^2+e_{\rm a}(\Delta z^2)) $,
where  $e_{\rm s}$ and $e_{\rm a}$ 
are the eigenvalues of $\bf{E}$, and 
$Z=P_{10}+P_{20}$ and $\Delta z=P_{10}-P_{20}$
are the symmetric and the antisymmetric superposition
of $u_{\rm mf}$ at the boundaries $\partial S_{i}$,
respectively. $L_y$ is the length of the boundary lines.
In fact,  the line densities $e_{\rm s}$ and $e_{\rm a}$
 correspond up to a factor to  the mean field energies
of the solutions of the mean-field equation~(\ref{mfeq}) with
 symmetric ($u_{\rm s}|_{\partial_{S_{1/2}}}=Z$)
 and antisymmetric  ($u_{\rm a}|_{\partial_{S_{1/2}}}=\pm \Delta z$)
  boundary conditions, respectively, 
and the general mean field solution is given by
 by $u_{\rm mf}=(u_{\rm s}+u_{\rm a})/2$.
 For $\lambda_c \to \infty$ the mean--field equation 
 reduces to the Laplace equation $\Delta u_{\rm mf}=0$
  between parallel lines which are in $x$-direction a distance $\tilde h$ apart.
 The antisymmetric and symmetric solutions read
 $u_{\rm a}=(2\Delta z/{\tilde h})\,x$  and $u_{\rm s}={\rm const.}$, respectively, with 
the corresponding line densities 
 $e_{\rm s}=0$
and $e_{\rm a}=4/{\tilde h}$.
A vanishing $e_{\rm s}$ signifies that a collective vertical shift
of the interface does not cost any energy.
This leads to the divergence of the integral over
$Z$ in the partition function $\mc{Z}_{\rm mf}$,
which physically is not interesting and can be neglected.
Applying the Derjaguin approximation (similar to  Eq.~(\ref{ffluc2})) to only the
antisymmetric mode yields the corresponding energy $E_{\rm a}$ for the
two circles a distance $h$ apart: 
%
 \begin{eqnarray}
 E_{\rm a}
 &\approx &
 -2 \gamma \int_0^{r_0} dy\,\frac{2(\Delta z)^2}{h+2r_0-2\sqrt{r_0^2-y^2}}
 \approx
 2  \pi\,\gamma\, (\Delta z)^2\sqrt{\frac{r_0}{h}} + \mc{O}(1)\;.
 \end{eqnarray}
Thus the $h$ dependent mean-field part of the free energy reads
$\mc{F}_{\rm mf}\simeq -k_{\rm B} T \ln\int d(\Delta z)\,\exp(-E_a/\,k_{\rm B} T) = 
(k_{\rm B}T/4)\ln(h/r_0) + {\rm const}$.
So, in the limit $h=d-2R \to 0$ the leading (divergent)
part of  
the effective mean-field force $F_{\rm mf}$
  is repulsive and reads
  \begin{equation}\label{derjmf}
   F_{\rm mf}(h\to 0) \approx \frac{k_{\rm B}T}{4\, h} \;.
  \end{equation}
It appears physically less obvious why the Derjaguin approximation could also be applied 
to higher boundary multipole moments $n$, especially if $h \alt R/n$. If one nevertheless
does so one finds that in this regime $h \alt R/n$ the mean--field force diverges slower
than $1/h$ whereas for $h \ll R/n$ the monopole behavior is recovered. This is in accordance
with our numerical results. 
\section{Intermediate distances: Numerical calculation}\label{numerics}
%
For intermediate distances $d\simeq r_0$ the fluctuation induced
force has to be calculated numerically.
We will do this as in the previous sections 
for the fluctuation and the  mean-field part separately.
For the fluctuation part, we shall apply a method which was introduced in Ref.~\cite{Bue04}.
The starting point is Eq.~(\ref{Zaux}) for the
partition function $Z_{\rm fluc}$.
Introducing an equidistant mesh with $N$ points
$\varphi_{ij}=(2\pi/N)j$, $0\le j < N$,
on the contact line
circles $\partial S_{i,\rm ref}$ converts 
the double integral in the exponent
to a double sum. Then the functional integrals
over the $\psi_i$ are replaced by ordinary
Gaussian integrals over the $\psi_i({\bf x}_i(\varphi_{ij}))$,
$\mc{D}\psi_i\simeq \prod_{j=0}^Nd\psi_i({\bf x}_i(\varphi_{ij}))$.
In the exponent, the $\psi_i({\bf x}_i(\varphi_{ij}))$ are
coupled by a matrix $\bf{G}$ with
elements $G_{ii'}^{jj'}=G(|{\bf x}_i(\varphi_{ij})-{\bf x}_{i'}(\varphi_{i'j'})|)$.
Performing the Gaussian integrals and 
disregarding divergent and $d$-independent terms immediately
leads to 
$\mc{F}_{\rm fluc}=(k_{\rm B}T/2)\ln\det({\bf G}_\infty^{-1}{\bf G}(d))$
for the fluctuation free energy.
Here, ${\bf G}_\infty\equiv \lim_{d\to\infty}{\bf G}(d)$.
It contains the self energy contributions and is
needed for the regularization of the free energy.
Deriving with respect to $d$, the Casimir
force can be written as 
\be\label{numforce}
F_{\rm fluc}(d)
=-\frac{k_{\rm B}T}{2}
\,{\rm tr}\left[
{\bf G}(d)^{-1}\partial_d{\bf G}(d)
\right]\;.
\ee
The advantage of the direct calculation of the
force is that Eq.~(\ref{numforce}) does not contain any 
  divergent parts which would require regularization,
thus easing the
numerical treatment considerably.
The determinant is computed by
using a standard LU decomposition \cite{press02}.
We find  good 
convergence of the numerical routine
also for small $d$ if $N\approx 5000$,
which, however, demands a computation time
of about 30 hours for each distance point on a standard PC.
As discussed in Sec.~\ref{sec:numres} and shown
in Figs.~\ref{figures}, we find  very good agreement between
analytical and numerical results.

The numerical calculation of the mean-field Casimir force
can be done very conveniently with the method described
in Sec.~\ref{sec:mf}.
In order to avoid complex numbers, we used 
$\sin$- and $\cos$-modes instead of Fourier-modes
for the numerical computations.
It is straightforward to rewrite the corresponding
equations in Sec.~\ref{sec:mf} by using their
real and imaginary parts.
Via a numerical evaluation of the integrals for the projection coefficients $a_{i,mn}$
in Eq.~(\ref{projcoeff}) and inversion of the linear system Eq.~(\ref{proj}) 
the  elements $E_{i,\rm self}^{mn}$
and $E_{12,\rm int}^{mn}$ of the matrix ${\bf E}$ are computed,
see Eqs.~(\ref{mfself}) and (\ref{mfint}). 
Calculating $\ln\det {\bf E}$
 then provides the free energy $\mc{F}_{\rm mf}$,
and the numerical derivative finally
the mean-field Casimir force $F_{\rm mf}$.
This scheme turns out to be very efficient
and provides results for $F_{\rm mf}$
within seconds.
Indeed, the numerical calculations show
that it is sufficient to consider
$n_{\rm max}\approx 20$ modes in the expansions
of the $u_i$ in Eq.~(\ref{superpos})
in order to achieve convergence of the results.


%
%
\section{Discussion
of results and of possible realizations}\label{sec:numres}
\begin{figure}

\begin{tabular}{cc}
\includegraphics[width=0.45\textwidth]{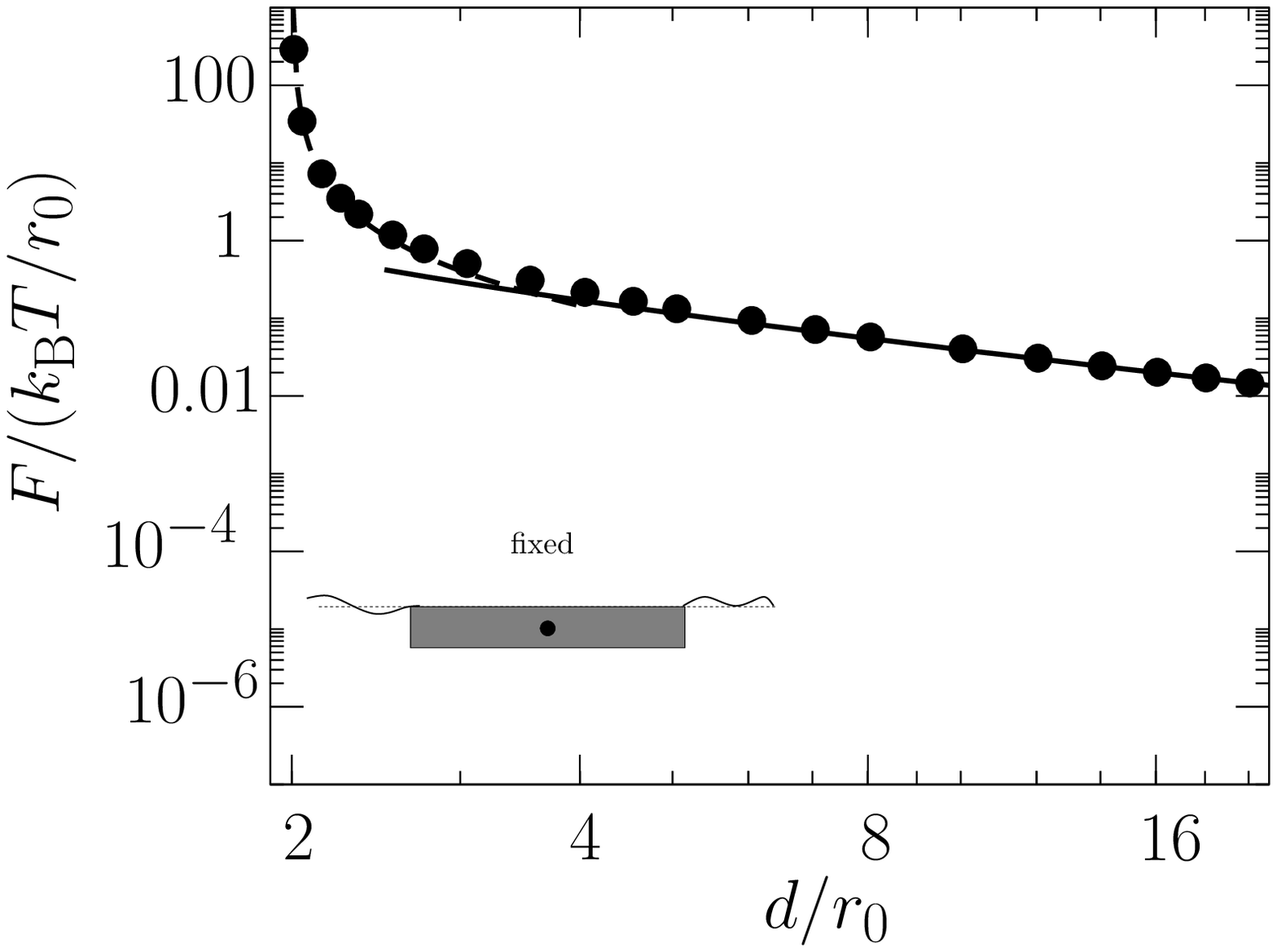}
\hspace*{0.5cm}
&
\includegraphics[width=0.45\textwidth]{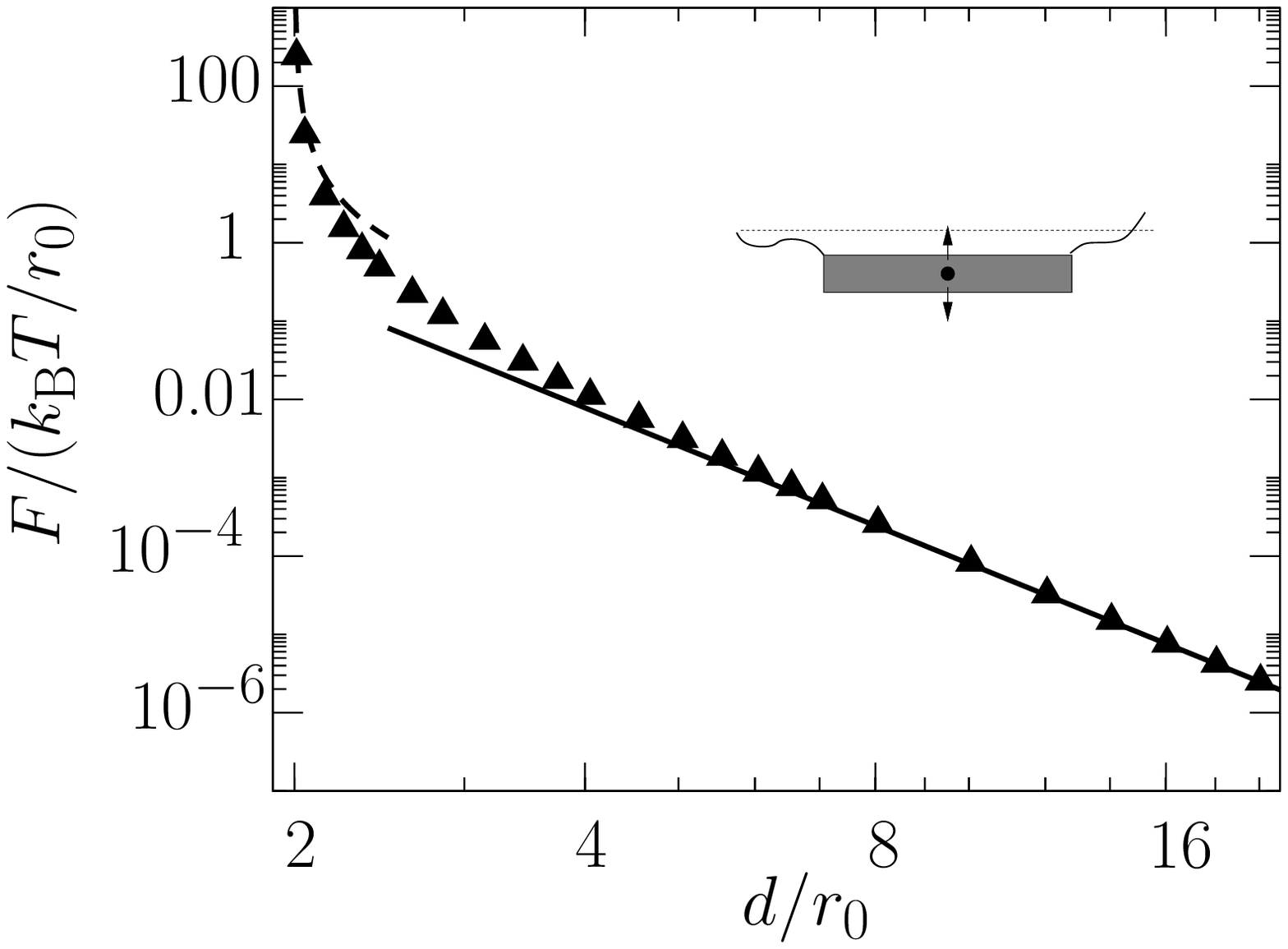}
\vspace*{1.5cm}
\\
\includegraphics[width=0.45\textwidth]{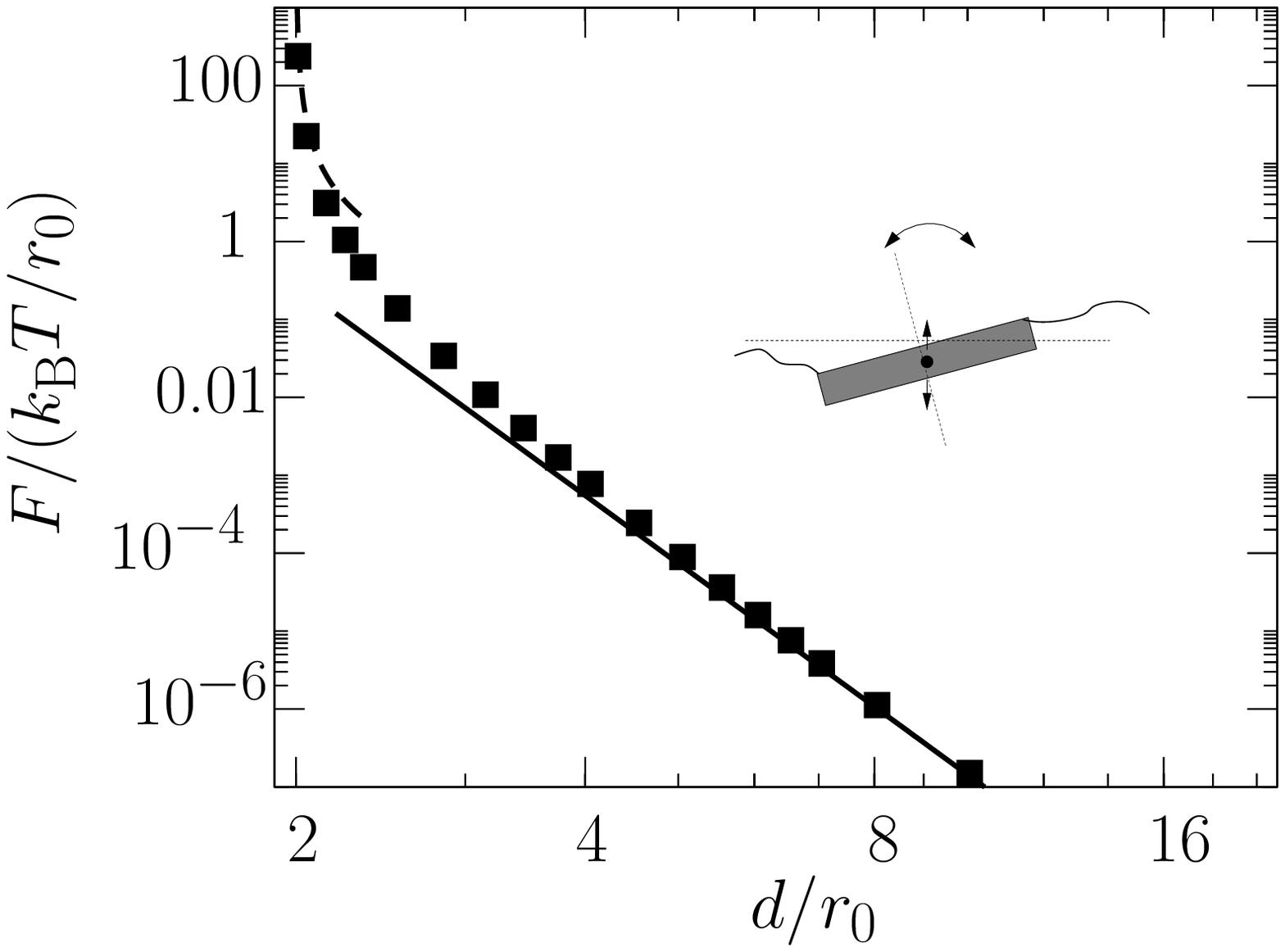}
\hspace*{0.5cm}
&
\includegraphics[width=0.45\textwidth]{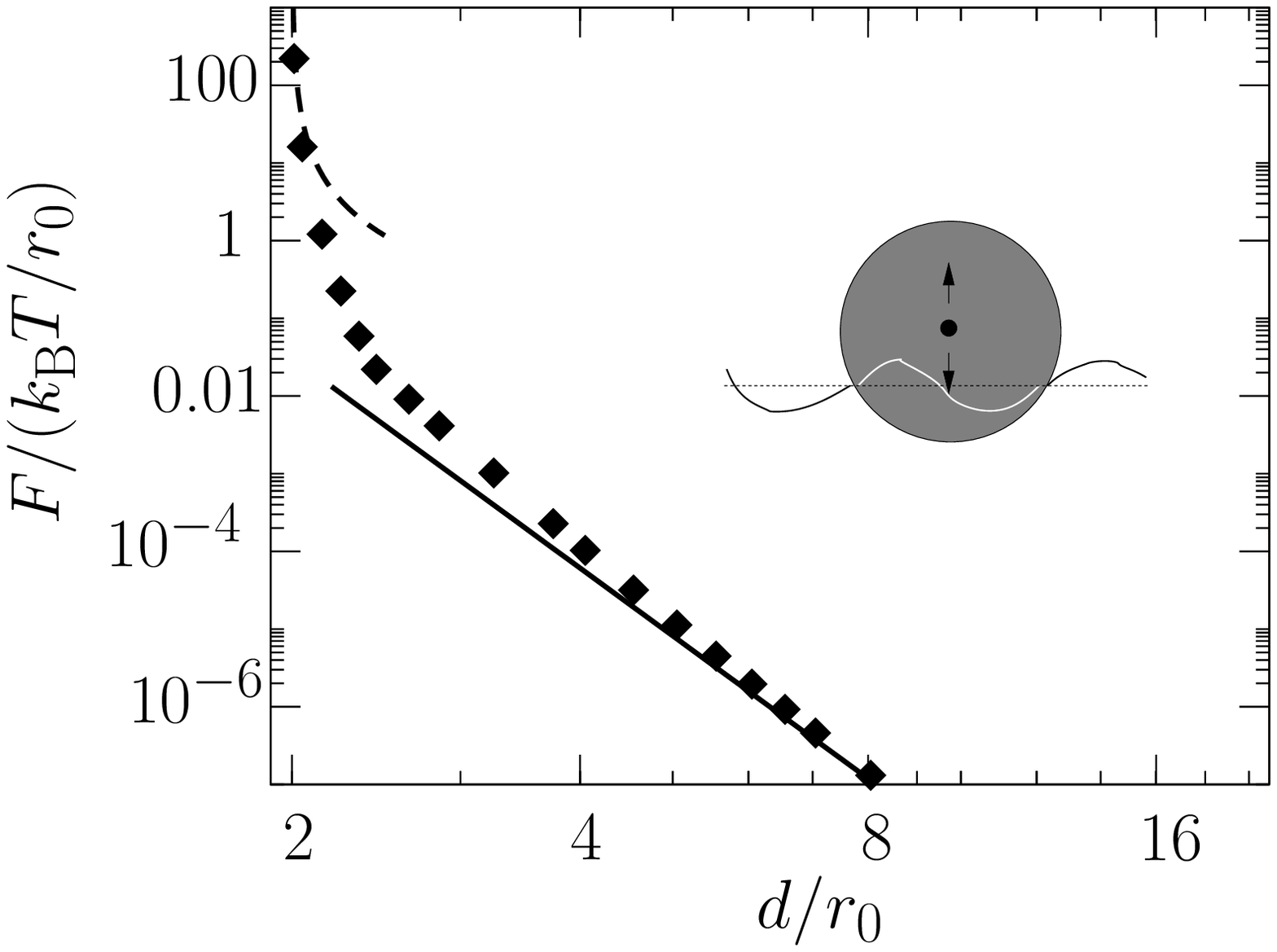}

\end{tabular}

\caption{Comparison of the numerical results for the Casimir force (symbols) with the 
analytical expressions
in the asymptotic ranges of large colloid separations $d \gg r_0$
(full line) and small
surface-to-surface distance $h=d-2r_0 \ll r_0$ (dashed line)
for the boundary condition
cases (B1), (B2), (B3), and (A),
respectively.
}
\label{figures}
\end{figure}

In Fig.~\ref{figures} we compare numerical results with the analytical expressions
for the Casimir force
 for the long-range asymptotics (Sec.~\ref{sec:anares})
and  for very small colloid-colloid
separations (Sec.~\ref{sec:derja}), respectively.

For all cases considered for the boundary conditions,
(B1)--(B3) and (A), we find a very good agreement with the 
analytical predictions,
Eqs.~(\ref{ffluc}) and
(\ref{lrmon})--(\ref{lrA})
in the long range regime for colloid separations $d \agt 5r_0$.
In the short-range regime,
the fluctuation induced force depends much less 
on the specific type of boundary conditions at the contact
line. In fact, in this regime, the divergence of 
fluctuation part force (which applies to all cases), $\sim -h^{-3/2}$,
is dominating the total force
in this regime and leads to a strong net attraction.
This may have an important influence on the effect
of colloid aggregation at interfaces as is discussed
in more detail in Ref.~\cite{Leh06}, especially with regard to
the van der Waals force. 
As can be seen from the plots in Fig.~\ref{figures},
in case (B1) (fixed colloids, no mean--field force) the Derjaguin approximation 
describes well the
numerical data also for intermediate colloid separations  $r_0/h=r_0/(d-2r_0)\le 5$.
In the plots for
cases (B2), (B3) and (A), however, we see that
the repulsive contribution from the mean-field part
showing a weaker divergence $\sim 1/h$
leads to a strong decrease of the total force for
$r_0/h=r_0/(d-2r_0)\ge 1$,
which, by increasing $d$, converges  to the power laws
governing the long-range asymptotics.
The rapid decrease of the total force actually
renders the Casimir force effectively short-ranged
for the cases (B3) and (A).
Note that because of this rapid decay $\sim -1/d^9$,
the competing
attractive and repulsive contributions from
the fluctuation and mean-field part are of almost equal size and
our numerical methods which are based on the addition
of these quantities are afflicted with numerical uncertanties 
and are not  able to provide reliable results for $d \agt 15r_0$.

In experimental realizations,
fixing the colloids as required in case (B1)
might, e.g.,
be realized by a laser tweezer. In such a setup, the
fixing is usually, of course, not exact, but 
means that the vertical movement of the
colloid centers is restricted by an external potential.
Including such a external potential for the
colloids in our model
in fact lowers the repulsive contribution from the mean-field
part (as compared to the unfixed case), and therefore 
leads to an increased long-ranged attractive 
total Casimir force dominated by
the leading term of the
fluctuation part, Eq.~(\ref{ffluc}).
The extent of this increase is controlled
by the strength of the external potential for
the colloids, 
as we will show in the next subsection.

\subsection{External potential on the colloids}\label{extpot}
In the effective Hamiltonian introduced
in Sec.~\ref{sec:model} 
we considered only free energy differences
resulting from the changes of interfacial areas
which are associated with fluctuations
around the flat reference configuration.
In this subsection we will extend this model
to external potentials $V_i(h_i)$ acting on the 
vertical position $h_i$ of the center of colloid $i$.
We will concentrate on the cases
of a constant external force $F_i {\bf e}_z$
and a harmonic potential for colloid $i$,
corresponding to $V_i(h_i)=-F_i h_i$
and $V_i(h_i)=\frac{1}{2}D_i\,(h_i-h_{0,i})^2$, respectively.
The vertical forces $F_i$ include, e.g., gravity,
whereas the harmonic potential can be realized
by a laser tweezer. 
So this is of particular interest for case (B1)
where  the colloid positions are fixed.

Then, the total effective Hamiltonian
of the two colloids adsorbed at the fluid interface
reads
\be\label{Hextpot}
\mc{H}=\mc{H}_{\rm cw}+\sum_i\left\{
\mc{H}_{{\rm b},i}+V_i(h_i)
\right\}\;.
\ee
Note, that the approximations performed in Sec.~\ref{sec:model}
in deriving the capillary wave and boundary Hamiltonian
remain valid here if the external force or the displacement
of the harmonic potential are small on the scale
set by the surface tension, i.e.,
$F_i \ll 2\pi\gamma r_0$ and $D_ih_0\ll 2\pi\gamma r_0$,
respectively.

The additional external potential has two implications which we
will discuss in the following. First, it leads
to a deformed equilibrium meniscus as compared to the flat
reference interface, which gives rise to a "classical" capillary
interaction between the colloids \cite{Oet05}.
Secondly, through the coupling of the
colloid position $h_i$ to the interface field $u$
in the boundary Hamiltonian $\mc{H}_{{\rm b},i}$,
the thermal movement of the colloids
in the potential $V_i$ can also influence 
the fluctuation induced ("{non} classical")
force between the colloids.

As described in detail in Ref~\cite{Oet05},
the equilibrium meniscus $u_{\rm eq}$
can be found by minimizing
the effective Hamiltonian $\mc{H}$
with respect to the colloid position $h_i$
and the interface height $u(x,y)$.
The equilibrium colloid height $h_{i,\rm eq}$
is determined from the condition
$\partial \mc{H}/\partial h_i=0$
and depends on both
the external potential and
the mean height of the
three phase contact line on the
colloid surfaces. 
%
The interface field $u_{\rm eq}$ fulfills
the Euler-Lagrange equation (\ref{mfeq}) of 
$\mc{H}_{\rm cw}$ with the boundary condition \cite{Oet05}
\bea\label{bceq}
\frac{\partial u_{\rm eq}(\bf{x}) }{\partial n_i}
&=&
\frac{u_{\rm eq}({\bf x})- h_{i,\rm eq}}{r_0} \;,
\eea
where $\partial/\partial n_i$ is the derivative in the
outward normal direction of $\partial S_{i,\rm ref}$.
Using the general form of Eqs.~(\ref{superpos}) and (\ref{superpos2})
 for $u_{\rm eq}$
and projecting the boundary conditions~(\ref{bceq})
 on $\partial S_{i,\rm ref}$
 onto $e^{im\varphi_i}$ 
as in Sec.~\ref{sec:mf}
leads again to a linear system for the expansion
coefficients $A_{im}$ similar to Eq.~(\ref{proj}).

For a constant external force,
we obtain
in the asymptotic range $r_0 \ll d \ll \lambda_c$
%
\bea\label{mensol}
u_{\rm eq}
&\simeq&
-\frac{F_1}{2\pi\gamma }\ln\left(\frac{\gamma_{\rm e}r_1}{2\lambda_c}\right)
-\frac{F_2r_0}{2\pi\gamma d }\left(\frac{r_0}{r_1}\right)\cos(\varphi_1)
-
\frac{F_2}{2\pi\gamma }\ln\left(\frac{\gamma_{\rm e}r_2}{2\lambda_c}\right)
-\frac{F_1 r_0}{2\pi\gamma d}\left(\frac{r_0}{r_2}\right)\cos(\varphi_2) 
\eea
for the leading terms in $r_0/d$ of the 
equilibrium meniscus $u_{\rm eq}$
(see Fig.~\ref{fig2} for the definitions of $r_i$ and $\varphi_i$).
The capillary interaction 
arising from the meniscus deformation
is given by \cite{Oet05}
\bea\label{Vmen}
V_{\rm men}(d)
&=&
\mc{H}(d)-\sum_i\mc{H}_{i,\infty}
\nonumber\\
&\approx&
\frac{F_1F_2}{2\pi\gamma}\ln\frac{\gamma_{\rm e}d}{2\lambda_c}
+\frac{F_1^2+F_2^2}{4\pi\gamma}\left(\frac{r_0}{d}\right)^2
\;.
\eea
Here, $\mc{H}_{i,\infty}$ is the effective energy
associated with a single colloid system.
The first term in the second line of Eq.~(\ref{Vmen})
is the well known logarithmic flotation force.
Note, that in the absence of electrostatic forces
the main contribution to the external force usually
is gravity which can be neglected for colloid radii
$\le 1 \mu{\rm m}$ as it is much smaller than the
thermal energy $k_{\rm B}T$. The second term
describes the leading behaviour of the
capillary interaction if there is an external force
only applied to one of the colloids
-- and it leads to a repulsive force between the colloids.

For  harmonic external potentials the equilibrium meniscus
and the capillary interaction potential 
is calculated in the same way but results in a
somewhat lengthy expression for $V_{\rm men}$.
Focusing onto the symmetric case of identical potentials
for both colloids, we find that
$V_{\rm men}$ vanishes in the limit $\lambda_c \to \infty$:
In the absence of gravity a parallel shift of the whole
interface does not cost any energy, and, therefore,
the response of the system on the presence of the
harmonic potentials on the colloids 
is a shift of the planar interface to a
new equilibrium position $u_{\rm eq}\equiv -h_{0}\,(=h_{0i}) $
-- without a meniscus deformation and, hence, without causing
a capillary interaction.
The leading term  for $1/\lambda_c$ small
is given by
\bea\label{capforceharm}
V_{\rm men}
&\simeq&
\frac{4\pi\gamma h_0^2 \ln(\gamma_{\rm e}d/2\lambda_c)}
{(2\pi\gamma/D-\ln(\gamma_{\rm e}r_0/2\lambda_c))
(2\pi\gamma/D-\ln(\gamma_{\rm e}r_0d/4\lambda_c^2))}\;.
\eea
If the minimum of the harmonic potentials on the colloids
are not identical, 
we find an additional term 
\bea\label{Vmenasy}
V_{\rm men}
&\stackrel{\lambda_c\to\infty}{\verylongrightarrow}&
\frac{\pi\gamma(h_{01}-h_{02})^2}{1+\pi\gamma(1/D_1+1/D_2)+\ln d/r_0}\;
\eea
which is not vanishing
in the limit $\lambda_c\to \infty$.


The influence of the external potential on the fluctuation induced
force is most conveniently discussed by splitting the
interfacial field into an equilibrium and a fluctuation
part, $u=u_{\rm eq}+\Delta u$, similar
to the splitting into a mean-field part $u_{\rm mf}$ and a fluctuation part $v$
described in Sec.~\ref{subs:mf},
and analogously $h_i=h_{i,\rm eq}+\Delta h_i$ for the
vertical position of the colloid centers. 
Note, however,
that $u_{\rm eq}$ fulfills the equilibrium 
boundary conditions~(\ref{bceq}), whereas
$u_{\rm mf}$ is the mean-field part of $\Delta u$,
i.e. $\Delta u = u_{\rm mf}+v$, which fullfills
the 
boundary conditions Eq.~(\ref{conlin})
corresponding to thermal fluctuations of the
contact line
around its equilibrium position.
Inserting this decomposition of the the
interfacial field $u$ into the extended Hamiltonian
Eq.~(\ref{Hextpot}) and performing some conversions
of the integrals by exploiting Eqs.~(\ref{mfeq}) and (\ref{bceq})
together with Gauss' theorem, we find a quite distinct
behaviour for the cases of a constant external force
and of a harmonic potential.
In the first case we can rewrite the effective Hamiltonian
in the form 
$\mc{H}[u_{\rm eq}+\Delta u]
=\mc{H}[u_{\rm eq}]+\mc{H}_{\rm cw}[\Delta u]+\mc{H}_{\rm b}[\Delta u,\Delta
h_i]$.
That means that the effective Hamiltonian 
relevant for the fluctuation induced force is independent
of the external force in this case.
The only effect of a constant external force, hence,
is the deformation of the equilibrium meniscus which leads
to the {\it classical} interaction described by Eq.~(\ref{Vmen}).
In the case of a harmonic external potential, however,
we find 
$\mc{H}[u_{\rm eq}+\Delta u]
=\mc{H}[u_{\rm eq}]+\mc{H}_{\rm cw}[\Delta u]+\mc{H}_{\rm b}[\Delta u,\Delta
h_i]+\sum_i (D_i/2)(\Delta h_i)^2$, i.e.
compared to the functional integral Eq.~(\ref{Z1})
we have to introduce an additional
harmonic potential term for the deviation $\Delta h_i$
of the colloid position from its equilibrium value $h_{i,\rm eq}$
in the expression for the partition function describing
the thermal fluctuations of the interface and the colloids.
The additional potential term
$V_i=(D_i/2)(\Delta h_i)^2$,
which is 
centered at $\Delta h_i=0$,
 can be
included in the boundary Hamiltonians $\mc{H}_{i,\rm b}$
and leads to additional terms in the self-energy parts of
Eq.~(\ref{mfself}) for the mean-field part of the partition function.
The integrals over the $\Delta h_i$ and the boundary multipoles
can be performed as before. 
The additional 
potentials $V_i$ leads to 
modifications of the
determinant of the matrix $\bf{E}$.
Because of the different form of integration measure
for the cases of a pinned und an unpinned contact line
(c.f. Sec.~\ref{sec:modelbc}),
these modifations differ for
the two types of boundary conditions.
 For the leading term
of the mean-field Casimir force
we can write
\bea\label{tweezerforce}
F_{\rm mf}
&\stackrel{\lambda_c\to\infty}{=}&
\frac{k_{\rm B}T}{2}
\frac{1}{d\ln d/r_0}
\frac{1}{1+\frac{\displaystyle o_1o_2 
\ln( d/r_0)}{\displaystyle o_1+o_2}}\;,
\eea
where $o_i=D_i/(\pi\gamma)$
for a pinned contact line and 
$o_i=2/(1+2\pi\gamma/D_i)$
for an unpinned contact line.
So the inclusion of the harmonic potential leads to a decreased mean-field contribution
to the fluctuation induced force as compared to Eq.~(\ref{fmf}) which
decreases with increasing strength $D_i$ of the potentials.
For a pinned contact line,
in this way the leading term of mean-field contribution
to the Casimir force can be switched on in a
controlled way by an external laser tweezer potential.
In the limit $D_i \to 0$ we recover 
for both cases
the result from Sec.~\ref{sec:mf}, Eq.~(\ref{fmf}),
which gives rise to the cancellation of the
leading terms with that of the fluctuation part.
This is expected because $D_i = 0$ corresponds
to the cases (A), (B2) or (B3) where height fluctuations of the colloids
are not suppressed by external potentials.
In the opposite limit $D_i \to \infty$, however,
we find $F_{\rm mf}\longrightarrow 0$ from Eq.~(\ref{tweezerforce})
for a pinned contact line. 
This corresponds to case (B1) 
of frozen colloid positions and a pinned contact line
and  where 
the full Casimir force is given by the fluctuation part Eq.~(\ref{ffluc}).
In the case of an unpinned contact line,
this effect is less pronounced.
Then the original form of the
mean-field Casimir force, Eq.~(\ref{fmf}),
is diminished by an additional factor
$1/(1+\ln d/r_0)$
in the limit $D_i \to \infty$.


For the linear external potential (constant force on the colloids)
-- also relevant for a laser tweezer
with relatively large displacement -- 
the only effect is an additional
capillary interaction to the fluctuation induced force.
The relative importance of these two contributions to
the colloid interaction is determined by their amplitudes
$F_{i}^2/\gamma$ and $k_{\rm B}T$, respectively.
For a harmonic external potential,
the relevance of the external potential on both
the classical capillary interaction (cf. Eqs.~(\ref{Vmen})--(\ref{Vmenasy}))
and on the
fluctuation induced force Eq.~(\ref{tweezerforce}),
is governed by the the ratio $\gamma/D_i$
of the surface tension and the stiffness, $D_i$,
of the harmonic potential.
For typical values for
the surface tension in the range of $\gamma \simeq 10$ mN/m
and a stiffness $D_i \simeq 10^{-3}-1$ mN/m \cite{Held06}
of the harmonic tweezer potential,
we find that the classical interaction can be
neglected compared to the fluctuation force (whose strength
is governed by $k_{\rm B}T$), but, also, the
effect of the external potential
on the Casimir force is marginal.
For steeper tweezer potentials or lower surface tensions
(c.f., e.g., Ref.~\cite{aarts04}),
such that $\gamma/D_i \leq 1$,
we find an increasing influence of the
the external potential.
Then the mean-field part $F_{\rm mf}$ of the fluctuation induced force
is considerably decreased which 
leads to a more attractive total Casimir force 
because the fluctuation part then will dominate $F_{\rm mf}$
(c.f. Sec.~\ref{sec:anares}).
On the other hand, in this regime also the classical capillary interaction
reaches the magnitude of the fluctuation induced force.

\begin{figure}
\psfrag{d}{\Large $d/r_0$}
\psfrag{Ftweezer}{\Large $F/(k_{\rm B}T/r_0)$}
\includegraphics[width=0.6\textwidth]{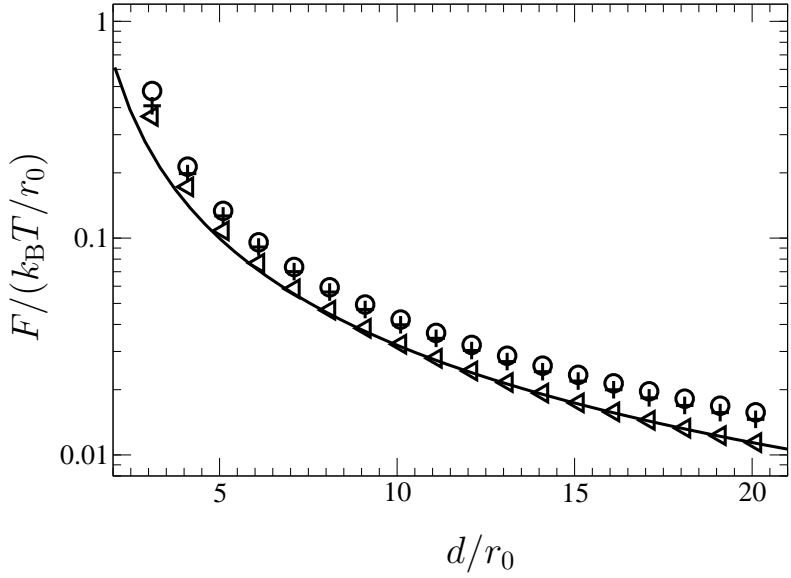}
\caption{
Numerical calculation of
mean-field part $F_{\rm mf}$ of the Casimir force
with harmonic external potential for the colloids 
for $D_i/\gamma=1$ (triangles, compared to the
analytical result Eq.~(\ref{tweezerforce}) solid line) and  
$D_i/\gamma=0.1$ (pluses)
in the case of pinned contact line.
The circles  show
the fluctuation part $-F_{\rm fluc}$
of the Casimir force,
which has to be added to $F_{\rm mf}$
to obtain the total force. As can be seen from the plot,
for increasing $D_i/\gamma$  
the fluctuation part becomes dominant.
}
\label{fignumtweezer}
\end{figure}
\section{Summary} 

The restrictions that two rotational symmetric colloids
trapped at a fluid interface impose
on the thermally excited interfacial fluctuations (capillary waves)
by their sheer presence lead to a thermal Casimir force.
This effective fluctuation force can be
calculated by a decomposition of the
interfacial partition function into a mean-field
and a fluctuation part
both numerically in the
whole range of colloid separations $d$ and analytically
for the asymptotic ranges of either small or large separations.
In the long-range limit, $d\gg r_0$, the resulting force
depends crucially on the boundary conditions at the
three-phase contact line on the colloids' surfaces.
There one observes an interesting interplay between
the attractive interaction from the interface fluctuations
and a repulsive interaction caused by the fluctuating
boundary conditions (mean-field part).
This results in a cancellation of the leading terms
up to a certain order in $1/d$, which is determined by
the specific model considered for the boundary conditions.
For freely fluctuating colloids -- either with a pinned (B3)
or unpinned (A) contact line --
the Casimir force is characterized by a fast decay
  $\propto -d^{-9}$.
For a pinned contact line, fixing colloidal degrees of freedom
leads to longer-ranged forces which are $\propto -d^{-5}$
if the orientation of the colloids is fixed and $\propto -1/d\ln d$
if both orientation and vertical position are fixed. 
This cancellation of the leading terms from the mean-field
and the fluctuation part can be understood in an alternative
approach to compute the partition function of the system
in which the analogy of the effective Hamiltonian
to electrostatics is exploited.
In this approach the fluctuation induced force
 can
be interpreted in terms of
an interaction between
auxiliary multipole moments defined
on the area enclosed by the  contact line of the colloids.
The various boundary conditions at the three phase contact line
translate into conditions for the
auxiliary multipoles. The asymptotics of the
fluctuation induced force for intermediate
colloid separations then is determined by
the interaction
of the leading non-vanishing
auxiliary multipoles.

In the opposite limit of a close colloid-colloid separation $h=d-2r_0 \ll
r_0$,
the effect of the boundary conditions is much less pronounced. Both the
mean-field and the fluctuation part diverge for $h \to 0$, but the
resulting  force is dominated by $F_{\rm fluc} \sim h^{-3/2}$
 (compared to $F_{\rm mf}\sim h^{-1}$), leading to a strong Casimir
 interaction in this regime (c.f. Fig. \ref{figures}).

For typical values in experimental situations, the
fluctuation force will dominate "classical" capillary forces
arising from meniscus deformations by an external potential.
Nevertheless,
such external potentials provide the possibility to tune
the fluctuation force directly
or to superimpose "classical" and fluctuation forces
by a sophisticated choice of the experimental setting.


\appendix
\section{The boundary Hamiltonian $\mc{H}_{\rm i,b}$}
\label{app:Hb}
From Eq.~(\ref{Hb}) we find that the boundary Hamiltonians
$\mc{H}_{\rm i,b}$ consists of three parts. In this appendix we
show how $\mc{H}_{\rm i,b}$ can be expressed in terms of the
boundary multipole moments $P_{im}$, leading to
the second order expansion given in  Eq.~(\ref{Hb}).
For the case (A) of spherical colloids and a fluctuating contact line all
areas $\Delta A_{\rm I/II}$ and $\Delta A_{\rm proj}$ are nonzero.
For the case (B3) of a pinned contact line (disks or Janus spheres),
$\Delta A_{\rm I/II}=0$ and the boundary Hamiltonians are fully determined by the change
in the projected meniscus area $\Delta A_{\rm proj}$. (For the remaining
cases (B1) and (B2) all area changes vanish and thus the boundary Hamiltonian is zero.) 
It is sufficient to determine
the area changes $\Delta A'_{\rm I/II}=0$ and $\Delta A'_{\rm proj}$ for a single
colloid; the total area changes are just given by a sum of these.

If the three phase contact is slowly varying 
without overhangs it can be written
as a function of polar angle $\varphi$, and for
 sperical colloids its distance to the $z$-axis
is given by
\be\label{tpcph} 
r_0(\varphi)
=\sqrt{R^2-(h_0+u(r_0(\varphi))^2}
=
\sqrt{r_0^2-2\,R\cos\theta\,u(r_0(\varphi))-u(r_0(\varphi))^2}
\,,
\ee
where $u(r_0(\varphi))$ is the height of the three phase contact line, 
and  $h_0=- R\cos \theta$ is the height of the
colloid center in the reference
configuration (c.f. Sec.~\ref{sec:model}).
For the second equality we have have used $r_0=R\sin\theta$.

Following App.~A in Ref.~\cite{Oet05}, we parametrize the
projection of the actual three phase contact line
onto the reference plane in terms of the polar angle $\varphi$ 
and write ($h$ is the change in vertical position of the
colloid center with respect to the reference configuration)
%
\bea\label{Hbapp1}
\gamma_{\rm I}\Delta A'_{\rm I}
+\gamma_{\rm II}\Delta A'_{\rm II}
&=&
\frac{\gamma}{2}\int_0^{2\pi}d\varphi\,
\left[
u(r_0(\varphi))-h
\right]^2
+\frac{\gamma}{2}\int_0^{2\pi}d\varphi\,
\left[
r_0^2(\varphi)-r_0^2
\right]
\nonumber\\
&\simeq&
\frac{\gamma}{2}\int_0^{2\pi}d\varphi\,
\left[
f-h
\right]^2
+\frac{\gamma}{2}\int_0^{2\pi}d\varphi\,
\left[
r_0^2(\varphi)-r_0^2
\right]\,,
\eea
where in the first term in second line we have replaced 
$u(r_0(\varphi)) \simeq u(r_0,\phi)\equiv f$, i.e. we have replaced
the contact line height by the meniscus height at the reference contact circle,
since this approximation produces only terms
which are at most of third order in the 
boundary multipoles.
The second contribution to $\mc{H}_{\rm b}$ stems
from the changes in the projected meniscus area
and can be written as
\bea\label{Hbapp2}
\gamma \Delta A'_{\rm proj}
&=&
\gamma
\int_0^{2\pi}d\varphi\int_{r_0(\varphi)}^{r_0}
dr\,r
=\frac{\gamma}{2}\int_0^{2\pi}d\varphi\,
\left[
r_0^2-r_0^2(\varphi)
\right]
\eea
In case (A), $\Delta A'_{\rm I,II}\neq 0$,
and both, Eq.~(\ref{Hbapp1})
and (\ref{Hbapp2}), are contributing to $\mc{H}_{i,\rm b}$,
leading to 
a cancellation  of the contrubution from the change in
the projected meniscus area of Eq.~(\ref{Hbapp2})
by the second term in Eq.~ (\ref{Hbapp2}).
Inserting the decomposition of $f(\varphi)$ from Eq.~(\ref{conlin})
(omitting the colloid label $i$) we find
\bea\label{Hbapp3}
\mc{H}_{\rm b}
&\simeq&
\frac{\gamma}{2}\int_0^{2\pi}d\varphi\,
\left[
f(\varphi)-h
\right]^2
\nonumber\\
&=&\frac{\pi\gamma}{2}\left[
2(P_{0}-h)^2
+4\sum_{m\ge 1}|P_{m}|^2
\right]\;.
\eea
For the case (B3) there are also tilt fluctuations 
of the vertical axis 
of the three phase
contact line circle 
which can be parametrized 
by  a boundary height according to
$f=P_{1}e^{i\varphi}+P_{-1}e^{-i\varphi}$,
the projection of the tilted circle onto
the reference plane $z=0$
is an ellipse, and we find for the boundary Hamiltonian
according to Eq.~(\ref{Hbapp2})
\bea\label{Hbapp5}
\mc{H}_{\rm b}
&\simeq&
\pi \gamma (|P_1|^2+|P_{-1}|^2)
\eea
Note, however, that we can write $\mc{H}_{{\rm b},i}$ as in Eq.~(\ref{Hb})
for all cases for the boundary conditions, (A) and (B1)--(B3), since
the integration measure $\mc{D}f_i$ for the contact line
is constructed such that only the actual terms for the respective cases
contribute, see Sec.~\ref{sec:modelbc}.

\section{Expansion of the Greens function}\label{app:schwing}
In this appendix we derive the multipole expansion of the Greens function 
$G(|{\bf x}|)\approx -(1/2\pi)\ln(\gamma_{\rm e}|{\bf x}|/2\lambda_c)$
between two ``charged'' (charges generating the auxiliary field $\psi_i$) and isolated 
regions ($\partial S_{i,\rm ref}$ in Sec.~\ref{sec:fluc} and  $S_{i,\rm ref}$ in 
Sec.~\ref{sec:Kardar}).
The method we apply is known from electrostatics and goes back
to Schwinger (c.f.~\cite{Schwing}) and is referred to as the Schwinger technique
in the literature. In doing so, we use the analogy of our
problem to two-dimensional electrostatics.
As starting point we use the fact that the logarithm 
is the generating function of the Gegenbauer polynomials.
Using $|{\bf r}-{\bf r'}|=\sqrt{r^2+r'^2-2rr'\cos(\varphi-\varphi')}$
and exploiting some properties of the Gegenbauer polynomials \cite{Abr74} we find
\be
\label{logex1}
\ln |{\bf r}-{\bf r'}|=\ln r 
+ \sum_{l\ge 1}\frac{1}{l}\left(\frac{r'}{r}\right)^l \cos(l\varphi-l\varphi')\;,
\ee
where we assumed $r'=|{\bf r}'|<r=|{\bf r}|$. Comparing Eq.~(\ref{logex1})
to the Taylor expansion of the logarithm, we find 
\be
\frac{(-{\bf r'}\cdot {\bf \nabla})^l}{l!}\,\ln r=
\left\{
\begin{array}{ll}
\ln r &\quad l=0 \\
-\frac{1}{2 l}\left(\frac{r'}{r}\right)^l\,\left(
e^{i(l\varphi-l\varphi')}+e^{-i(l\varphi-l\varphi')}
\right)
&\quad l>0
\end{array}
\right.\;.
\ee
On the other hand, introducing $\xi_\pm=\partial_x \pm i\partial_y$
and using $\xi_+\xi_-\ln r=\xi_-\xi_+\ln r=0$,
we can write
\bea\label{logex2}
\frac{(-{\bf r'}\cdot {\bf \nabla})^l}{l!}\,\ln r
=\frac{(-1)^l}{l!}\left(\frac{r'}{2}\right)^l
\left(
e^{-il\varphi'}\xi_+^l + e^{il\varphi'}\xi_-^l
\right)\,\ln r
\eea
Identifying Eqs.~(\ref{logex1}) and (\ref{logex2})
we obtain
\be\label{xiex} 
\xi_\pm^l\,\ln r
=-(-2)^{l-1}(l-1)!\frac{e^{\pm il\varphi}}{r^l}
\ee
With Eq.~(\ref{xiex}) at hand and
for  ${\bf x}_{1}={\bf r}_1$ and ${\bf x}_2={\bf d}+ {\bf r}_{2}$ 
residing on different circles
we obtain for the Greens function $G(|{\bf x}_2-{\bf x}_1|)$
\bea\label{greensmulti}
-\frac{1}{2\pi}\ln\left(\frac{\gamma_{\rm e}|{\bf d}+{\bf r}_2-{\bf r}_1|}
{2\lambda_c}\right)
&=&
-\frac{1}{2\pi}\ln\left(\frac{\gamma_{\rm e} d}{2\lambda_c}\right)
+\frac{1}{2\pi}\sum_{l_1,l_2 =0  \atop l_1+l_2 \ge 1}
\frac{(-{\bf r}_1\cdot {\bf \nabla})^{l_1}}{l_1!}
\frac{({\bf r}_2\cdot {\bf \nabla})^{l_2}}{l_2!}\,\ln r
\nonumber\\
&=&
-\frac{1}{2\pi}\ln\left(\frac{\gamma_{\rm e} d}{2\lambda_c}\right)
\nonumber\\&&+
\frac{1}{2\pi}\sum_{l_1,l_2 =0  \atop l_1+l_2 \ge 1}
\frac{(-r_1)^{l_1}r_2^{l_2}}{l_1!l_2!2^{l_1+l_2}}
\left[
e^{-i(l_1\varphi_1+l_2\varphi_2)}\xi_+^{l_1+l_2} 
+ e^{i(l_1\varphi_1+l_2\varphi_2)}\xi_-^{l_1+l_2}
\right]\,\ln r
\nonumber\\
&=&
-\frac{1}{2\pi}\ln\left(\frac{\gamma_{\rm e} d}{2\lambda_c}\right)
\nonumber\\&&+
\frac{1}{2\pi}\sum_{l_1,l_2 =0  \atop l_1+l_2 \ge 1}
\frac{(-1)^{l_1}}{l_1+l_2}{l_1+l_2 \choose l_1}
\frac{r_1^{l_1}r_2^{l_2}}{2d^{l_1+l_2}}\,
\left[
e^{-i(l_1\varphi_1+l_2\varphi_2)} 
+ e^{i(l_1\varphi_1+l_2\varphi_2)}
\right]\;.
\eea
In Sec.~\ref{sec:fluc} we also need the multipole expansion of 
$G(|{\bf x}_2-{\bf x}_1|)$ for ${\bf x}_i$ residing on the circumference
$\partial S_{i,\rm ref}$ of the same circle
in order to calculate the elements of the self-energy matrix, Eq.~(\ref{flucse}).
Using Eq.~(\ref{logex1}), we obtain with
$|{\bf x_1}-{\bf x_2}|=r_0\sqrt{2-2\cos(\varphi_1-\varphi_2)}$ 
\bea\label{ftse}
G(|{\bf x}_1-{\bf x}_2|)
&\simeq & 
-\frac{1}{2\pi}\ln\left(\frac{\gamma_{\rm e} r_0}{2\lambda_c}\right)
-\frac{1}{2\pi}\ln[1-\cos(\varphi_2-\varphi_1)])
\nonumber\\
&=&
-\frac{1}{2\pi}\ln\left(\frac{\gamma_{\rm e} r_0}{2\lambda_c}\right)
+\frac{1}{2\pi}\sum_{l\ge 1}\frac{1}{ 2l}
[e^{i l\varphi_1- il\varphi_2}+e^{-i l\varphi_1+i l\varphi_2}]\,.
\eea
Here the prerequisite $r'<r$ of Eq.~(\ref{logex1}) is not fulfilled,
and the series in Eq.~(\ref{ftse}) is not convergent. It has to be
understood
in a formal sense, since it only provides the Fourier coefficients
for a finite number of modes which actually contribute to the effective 
interaction in the long-range regime (see Sec.~\ref{sec:fluc} for details).

\section{Kardar's method: calculation of the self-energy}\label{app:se}
The self--energy part, Eq.~(\ref{sekard}), is evaluated  quite similarly as in 
Ref.~\cite{Gol96}. We eliminate the $\delta$ functions by introducing
conjugate multipole moments $\widetilde{\Psi}_{im}$ of the auxiliary fields:
\begin{equation}
 \delta
\left(\Psi_{im} -\int_{S_{i,\rm ref}}d^2x\,(r/r_0)^{|m|}e^{-im\varphi}\psi({\bf x})\right)
 =  \int d\widetilde{\Psi}_{im} \exp\left( i\widetilde{\Psi}_{im}
 \left[\Psi_{im} -\int_{S_{i,\rm ref}}d^2x\,(r/r_0)^{|m|}
e^{-im\varphi}\psi({\bf x})\right]\right)\;. 
\end{equation} 
This brings $\mc{Z}_{\rm i,self}=\exp\{ -k_{\rm B}T/(2\gamma)\,\mc{H}_{i,\rm self}\}$
into the form
\bea\label{sekardapp}
\mc{Z}_{\rm i,self}
&=& \int \prod_m d\widetilde{\Psi}_{im}
\int \mc{D}\psi_{i}\,\exp\left\{
-\frac{k_{\rm B}T}{2\gamma}
\int_{S_{i,\rm ref}}d^2x \int_{S_{i,\rm ref}}d^2x'\,
\psi_i({\bf x})\,G(|{\bf x}-{\bf x'}|)\,\psi_i({\bf x})
\right.\nonumber\\&&\qquad\left.
-i\int_{S_{i,\rm ref}}d^2x\,\psi_i({\bf x})\,\sum_{m}
\left(\frac{r_i}{r_0}\right)^{|m|}
(P_{im}+\widetilde{\Psi}_{im})e^{i m\varphi_i} +i\sum_{m=-\infty}^\infty
(P_{im}+ \widetilde{\Psi}_{im})\Psi_{im}
\right\}
\;.
\eea
The functional integral $\int\mc{D}\psi_i$ in Eq.~(\ref{sekardapp})
can be converted into a functional integral over a constrained height field $h({\bf x})$;
this corresponds to a reversion of the step from Eq.~(\ref{Zkard1})
to Eq.~(\ref{Zkard2}),
\bea\label{sekard1}
\mc{Z}_{i,\rm self}&=&\int \prod_m d\widetilde{\Psi}_{im}\,
 \exp\{i \sum_{m=-\infty}^\infty 
(P_{im}+\widetilde{\Psi}_{im})\Psi_{im}\}
\int\mc{D}h\,
\prod_{{\bf x}_i \in S_{i,\rm ref}}
\delta\left(
h({\bf x}_i)-
\sum_{m}
\left(\frac{r_i}{r_0}\right)^{|m|}
(P_{im}+\widetilde{\Psi}_{im})e^{i m\varphi_i}
\right)
\nonumber\\&&\qquad\qquad
\times\exp\left\{
-\frac{\gamma}{2k_{\rm B}T}\int d^2x\,\left[
(\nabla h)^2+\frac{h^2}{\lambda_c^2}
\right]
\right\}
\;.
\eea
In Eq. (\ref{sekard1}), the $\delta$-functions describe
the pinning of the field $h$ in the region $S_{i,\rm ref}$. 
This contribution can be evaluated directly, such that
the remaining functional integral reads 
\bea\label{sekard2}
\mc{Z}_{\rm i,self}
&\stackrel{\lambda_c\to\infty}{\approx}&\int \prod_m d\widetilde{\Psi}_{im}
\exp\left\{
 -\frac{2\pi\gamma}{k_{\rm B}T} \sum_{m\ge 1} m
  |P_{im}+\widetilde{\Psi}_{im}|^2 +i \sum_m(P_{im}+\widetilde{\Psi}_{im})\Psi_{im} 
\right\}
\nonumber\\&&\times
\int\mc{D}h\,
\prod_{{\bf x}_i \in \partial S_{i,\rm ref}}
\delta\left(h({\bf x}_i)-
\sum_{m}
(P_{im}+\widetilde{\Psi}_{im})e^{i m\varphi_i}
\right)
\nonumber\\&&\qquad\qquad
\times \exp\left\{
-\frac{\gamma}{2k_{\rm B}T}\int_{\mathbb{R}^2\setminus S_{i,\rm ref}} d^2x\,\left[
(\nabla h)^2+\frac{h^2}{\lambda_2^2}
\right]
\right\}
\;,
\eea
where the $\delta$-functions fix $h({\bf x})$ at the
boundaries $\partial S_{i,\rm ref}$ of the integration domain.
Splitting the auxiliary field into two parts as in Eq.~(\ref{mffluc}),
 $h=h_0+ h_1$ , where 
$(-\Delta+\lambda_c^{-2})\,h_0=0$ with the boundary conditions
$h_0({\bf x}_i)|_{\partial S_{i,\rm ref}}=\sum_{m=-\infty}^\infty
(P_{im}+\widetilde{\Psi}_{im})e^{i m\varphi_i}$
and $ h_1({\bf x}_i)|_{\partial S_{i,\rm ref}}=0$, and applying Gauss' theorem
to the integral in the exponent of Eq.~(\ref{sekard2}) 
 leads to
\bea\label{sekard3}
\mc{Z}_{\rm i,self}
&=&\int \prod_m d\widetilde{\Psi}_{im}
\exp\left\{
 -\frac{2\pi\gamma}{k_{\rm B}T} \sum_{m \ge 1} |m|
  |P_{im}+\widetilde{\Psi}_{im}|^2 + i\sum_{m=-\infty}^\infty
(P_{im}+\widetilde{\Psi}_{im})\Psi_{im}
\right\}
\nonumber\\&&\times
\exp\left\{
- \frac{\gamma}{2k_{\rm B}T} 
\oint_{\partial S_{i,\rm ref}} d{\bf x}\,
h_0({\bf x})\nabla h_0({\bf x})
\right\}
\nonumber\\&&\times
\int\mc{D} h_1\,
\prod_{{\bf x} \in \partial S_{i,\rm ref}}
\delta\left( h_1({\bf x})\right)
\exp\left\{
-\frac{k_{\rm B}T}{2\gamma}\int_{\mathbb{R}^2\setminus S_{i,\rm ref}} d^2x\,\left[
(\nabla  h_1)^2+\frac{ h_1^2}{\lambda_c^2}
\right]
\right\}
\;.
\eea
The functional integral over $ h_1$ only yields a
constant factor independent of any multipole moment,
which will be disregarded below. 
To compute the line integral in Eq.~(\ref{sekard3}), we write 
the general solution of the differential equation for $h_0$
in $\mathbb{R}^2\setminus S_{i,\rm ref}$
as (c.f. Sec.~\ref{sec:mf})
$h_0({\bf x})=\sum_{m}(K_m(r/\lambda_c)/K_m(r_0/\lambda_c))
C_m e^{i m\varphi}$.
By comparison to the boundary conditions 
 the coefficients are determined straightforwardly
 as
$C_m=P_{im}+\widetilde{\Psi}_{im}$.
Then, the line integral evaluates to
$2\pi|P_{im}+\widetilde{\Psi}_{im}|^2\,f(m)$ with
$f(m)= |m|$ $(|m| \ge 1)$
and $f(0)=-1/(\ln(\gamma_{\rm e} r_0/\,2\lambda_c))$,
such that the self energy part reads
\bea\label{sekard4}
\mc{Z}_{i,\rm self}
&=&\int \prod_m d\widetilde{\Psi}_{im}
\exp\left\{
 -\frac{4\pi\gamma}{k_{\rm B}T} \sum_{m \ge 0} \frac{f(m)}{1+\delta_{m0}} 
  |P_{im}+\widetilde{\Psi}_{im}|^2  +i\sum_{m=-\infty}^\infty
(P_{im}+\widetilde{\Psi}_{im}) \Psi_{im}
\right\}\;.
\eea
and thus $\mc{H}_{i,\rm self}$ is given by (up to unimportant additive constants)
\begin{equation}
 \mc{H}_{i,\rm self} = \sum_{m\ge 1} |\Psi_{im}|^2 \frac{1}{2\pi\,f(m)} \;.
\end{equation}

\end{document}